\pdfoutput=1

\documentclass[a4paper,10pt]{article}

\usepackage{contribution}

\newcommand{\weblink}[2][]{%
    \ifthenelse{\equal{#1}{}}%
    {\textnormal{\url{#2}}}%
    {\textnormal{\href{#2}{#1}}}%
}

\def\beq{\begin{equation}}
\def\eeq#1{\label{#1}\end{equation}}
\def\eeqn{\end{equation}}

\def\beqa{\begin{eqnarray}}
\def\eeqa#1{\label{#1}\end{eqnarray}}
\def\eeqan{\end{eqnarray}}

\let\bar=\overbar

\def\etal{{\it et al.}}

\def\Dslash{\not{\hbox{\kern-4pt $D$}}}
\def\dslash{\not{\hbox{\kern-2pt $\del$}}}

\def\ee{e^+e^-}

\def\msb{{\bar{\ssstyle M \kern -1pt S}}}

\newcommand{\contribution}[7][]{%
  \clearpage
  \thispagestyle{plain}

  \ifthenelse{\equal{#1}{}}
  {\hypersetup{pdftitle={#2}}}
  {\hypersetup{pdftitle={#1}}}
  \hypersetup{pdfauthor={{#3} {#4}}}
  {\centering\normalfont\LARGE\bfseries\sffamily #2 \par\nobreak}
  \lhead{}
  \chead{%
    \textit{\footnotesize XXIInd International Workshop ``High-Energy Physics and Quantum Field Theory'', 
June 24 -- July 1, 2015, Samara, Russia}%
  }
  \rhead{}
  \bigskip
  \begin{center}
    {#3} {#4}\ifthenelse{\equal{#6}{}}{}{\footnote{\weblink[#6]{mailto:#6}}}
    \ifthenelse{\equal{#7}{}}{}{#7} \\
    \textit{#5}
  \end{center}
  \bigskip
}

\renewcommand{\abstract}[1]{%
  \begin{center}
    \begin{minipage}{0.85\textwidth}
      \begin{footnotesize}
        #1
      \end{footnotesize}
    \end{minipage}
  \end{center}
  \bigskip
}

\begin{document} 

%
%
%
%
%
%
{  


\newcommand{\leplep}{\ell^{+}\ell^{-}}
\newcommand{\jp}{J/\psi}
\newcommand{\ec}{\eta_{c}}
\newcommand{\ecp}{\eta_{c}(2S)}
\newcommand{\psip}{\psi^{\prime}}
\newcommand{\etap}{\eta^{\prime}}
\newcommand{\chicone}{\chi_{c1}}
\newcommand{\onedtwo}{\eta_{c2}}
\newcommand{\threedtwo}{\psi_{2}(1{\rm D})}
\newcommand{\mumu}{\mu^{+}\mu^{-}}
\newcommand{\pipi}{\pi^{+}\pi^{-}}
\newcommand{\kk}{K^{+}K^{-}}
\newcommand{\pim}{\pi^{-}}
\newcommand{\pip}{\pi^{+}}
\newcommand{\piz}{\pi^{0}}
\newcommand{\ppbar}{p\bar{p}}
\newcommand{\nnbar}{n\bar{n}}
\newcommand{\NNbar}{N\bar{N}}
\newcommand{\pb}{\bar{p}}
\newcommand{\nb}{\bar{n}}
\newcommand{\uubar}{u\bar{u}}
\newcommand{\ubar}{\bar{u}}
\newcommand{\ddbar}{d\bar{d}}
\newcommand{\dbar}{\bar{d}}
\newcommand{\DDbar}{D\bar{D}}
\newcommand{\ssbar}{s\bar{s}}
\newcommand{\ccbar}{c\bar{c}}
\newcommand{\cbar}{\bar{c}}
\newcommand{\bbbar}{b\bar{b}}
\newcommand{\BBbar}{B\bar{B}}
\newcommand{\bbar}{B\bar{B}}
\newcommand{\qqbar}{q\bar{q}}
\newcommand{\qbar}{\bar{q}}
\newcommand{\QQbar}{Q\bar{Q}}
\newcommand{\kpi}{K^-\pi^{+}}
\newcommand{\ks}{K_{s}}
\newcommand{\kstr}{K^{*0}}
\newcommand{\chip}{\chi^{\prime}}
\newcommand{\B}{B^{0}}
\newcommand{\Bpl}{B^{+}}
\newcommand{\kpl}{K^{+}}
\newcommand{\kmi}{K^{-}}
\newcommand{\fb}{f b^{-1}}
\newcommand{\Mbc}{M_{\rm bc}}
\newcommand{\DE}{\Delta E}
\newcommand{\ra}{\rightarrow}
\newcommand{\rt}{\rightarrow}
\newcommand{\ETAL}{\em et al.}
\newcommand{\jpsi}{J/\psi}
\newcommand{\ximm}{\Xi^{--}}
\newcommand{\xic}{\Xi^0_c(2470)}
\newcommand{\xim}{\Xi^{-}}
\newcommand{\ximmbar}{\bar{\Xi}^{++}}
\newcommand{\ximbar}{\bar{\Xi}^{+}}
\newcommand{\xistr}{\Xi^{*0}(1530)}
\newcommand{\xistrbar}{\bar{\Xi}^{*}(1530)}
\newcommand{\lam}{\Lambda^{0}}
\newcommand{\lm}{\Lambda}
\newcommand{\lambar}{\bar{\Lambda}^{0}}
\newcommand{\lmbar}{\bar{\Lambda}}
\newcommand{\lmb}{\bar{\Lambda}}
\newcommand{\ups}{\Upsilon(1S)}
\newcommand{\upstwos}{\Upsilon(2S)}
\newcommand{\upss}{\Upsilon(1,2S)}
\newcommand{\lmppi}{\Lambda p \pi^-}
\newcommand{\lmbpbpi}{\bar{\Lambda} \bar{p} \pi^+}
\newcommand{\yones}{\Upsilon(1S)}
\newcommand{\ytwos}{\Upsilon(2S)}
\newcommand{\yonetwos}{\Upsilon(1,2S)}
\newcommand{\yns}{\Upsilon(nS)}
\newcommand{\twog}{\gamma\gamma}
\newcommand{\babar}{BaBar}
\newcommand{\DZero}{D0}

%

\contribution[History of Belle and some highlights]  
{History of Belle and some of its lesser known highlights}  
{Stephen Lars}{Olsen}  
{Center for Underground Physics, Institute for Basic Science \\
Yuseong-gu, Daejeon  Korea}  
{solsensnu@gmail.com}  
{}  
%


\abstract{
I report on the early history of Belle, which was almost entirely focused on testing
the Kobayashi Maskawa mechanism for $CP$ violation that predicted large matter-antimatter
asymmetries in certain $B$ meson decay modes.  Results reported by both BaBar and Belle
in the summer of 2001 verified the Kobayashi Maskawa idea and led to their Nobel
prizes in 2008.  In addition to studies of $CP$ violation, Belle (and BaBar) reported a
large number of important results on a wide variety of other subjects, many of which that
had nothing to do with $B$ mesons.  In this talk I cover three (of many) subjects
where Belle measurements have had a significant impact on specific sub-fields of hadron
physics but are not generally well know.  These include: the discovery of an anomalously
large cross sections for double charmonium production in continuum $\ee$ annihilation;
sensitive probes of the structure of the low-mass scalar mesons; and first measurements
of the Collins spin fragmentation function.
}


\section{Introduction}

The organizers of this meeting asked me to give a talk with the title ``Best Belle
results and history of Belle collaboration.'' A talk about the history of Belle is no
problem.  The intial motivation of the Belle/KEKB project and essentially all of its
early work was the study of $CP$-violation in the $B$-meson sector, and the work
done in this area certainly ranks among Belle's ``best results.'' However, while the planing
and first few year's of operation were almost completely focused on $CP$-violation
measurements, the collaboration subsequently branched out and studied a wide range of physics
subjects that included many unexpected discoveries.  To the people involved, each of
these rank among the ``best Belle results,'' and I could not argue with them.  So instead
of even attempting to identify ``best results,'' I decided to confine myself to
reporting on Belle's early work on $CP$ violation, which covers the early ``history of
the Belle collaboration,'' and then discuss a few other results that seem to be
not very widely known but have had a huge impact on the specialized areas of physics
that they address.  First, some history and Belle's early $CPV$ measurements:

\section{Belle and $\mathbf{CP}$ violation}
The Belle experiment traces its roots to the 1964 discovery that the long-lived neutral
kaon ($K_L$) is not a $CP$ eigenstate, as evidenced by a small but non-zero branching
fraction to $\pipi$: ${\mathcal B}(K_L\rt\pipi)\simeq 2\times 10^{-3}$, which 
demonstrated that $CP$ is violated, probably by the weak interactions~\cite{christenson64}.  
This inspired Sakharov's classic 1967 paper~\cite{sakharov67} that pointed out that
$CP$-violation ($CPV$) is an essential ingredient for explaining the baryon asymmetry of the
universe; {\it i.e.}, how a matter-antimatter-symmetric condition that prevailed right after the Big Bang,
evolved into today's matter-dominated universe (see Fig.~\ref{fig:sakharov}).

\begin{figure}[h!]

\begin{center}
\includegraphics[width=14cm]{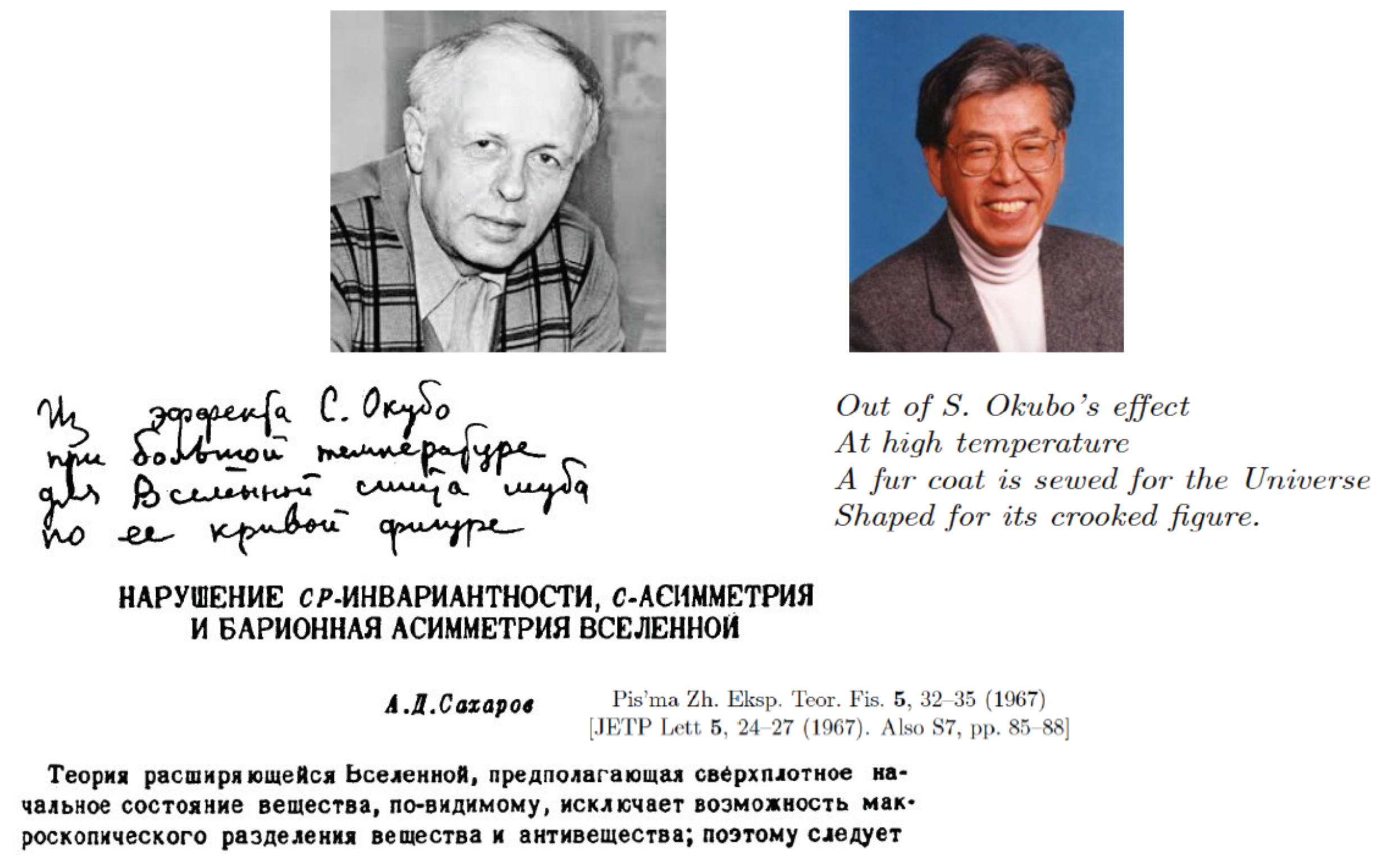}
\caption{\footnotesize
When Sakharov completed  his famous 1967 paper on ``Violation of $CP$ invariance, $C$ asymmetry,
and baryon asymmetry of the universe,'' he gave Lev Okun a preprint with a small poem handwritten
on it that identifies $CPV$ as ``S. Okubo's effect.'' This refers to a 1958 paper by Okubo that
first pointed out that while $CPT$ invariance requires particle and antiparticle lifetimes to be
equal, $CP$ violations would allow partial lifetimes to be different~\cite{okubo58}.}
\label{fig:sakharov}
\end{center}
\end{figure} 

Incorporating $CPV$ into the Standard Model (SM) while preserving $CPT$ was not very easy.  Wolfenstein
proposed a mechanism that expanded the SM by adding a new, $\Delta S=2$ ``superweak'' interaction
that produced a $CP$-violating non-diagonal contribution to the neutral kaon mass matrix and
nothing else~\cite{wolfenstein64}.  However, the superweak interaction was ruled out by the
observation of direct $CPV$ decays of neutral kaons by the NA31 experiment at CERN~\cite{na31}
and, later, the KTeV experiment at Fermilab~\cite{ktev}.  

To incorporate $CPV$ into the SM proper, one needs an amplitude that has a complex phase angle $\phi_{CP}$
that has opposite signs for particle and antiparticle processes.  Since measureable processes are
proportional to the absolute value squared of the amplitude, this $CPV$ phase is unmeasureable
unless it interferes with another process that has a non-zero strong, or common phase $\phi_0$,
that has the same sign for particle and antiparticle processes~\cite{pakvasa83}.  This
is illustrated in Fig.~\ref{fig:cp-phase}a, where a $CP$ violating process $X^0\rt\pipi$
($\bar{X}^0\rt\pipi$) has a complex amplitude $A=|A|\exp(i\phi_{CP})$ ($\bar{A}=|A|\exp(-i\phi_{CP}$)).
Differences in the decay rates can be be measured if the $CP$-violating amplitude interferes with a
$CP$-conserving amplitude for the same process $C=|C|\exp(i\phi_0)=\bar{C}$.  In that case the 
$X^0\rt\pipi$ and $\bar{X}^0\rt\pipi$ rates differ by a term proportional to
$2|A||C|\sin\phi_0\sin\phi_{CP}$; note that this interference term is zero if $\phi_0 =0$.  

\begin{figure}[htb]
\begin{minipage}[t]{70mm}
  \includegraphics[height=1.1\textwidth,width=0.9\textwidth]{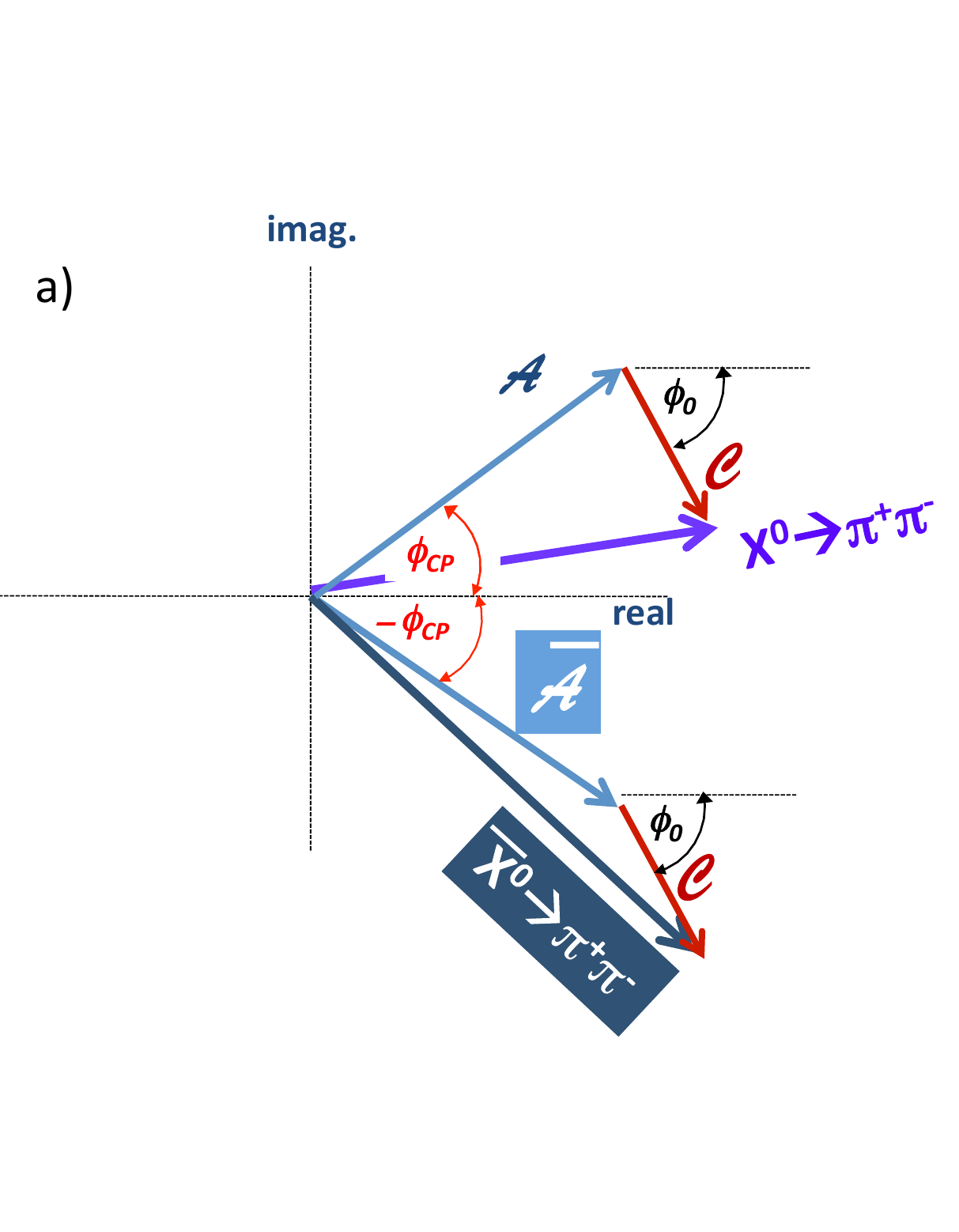}
\end{minipage}
\begin{minipage}[t]{96mm}
  \includegraphics[height=0.7\textwidth,width=1.0\textwidth]{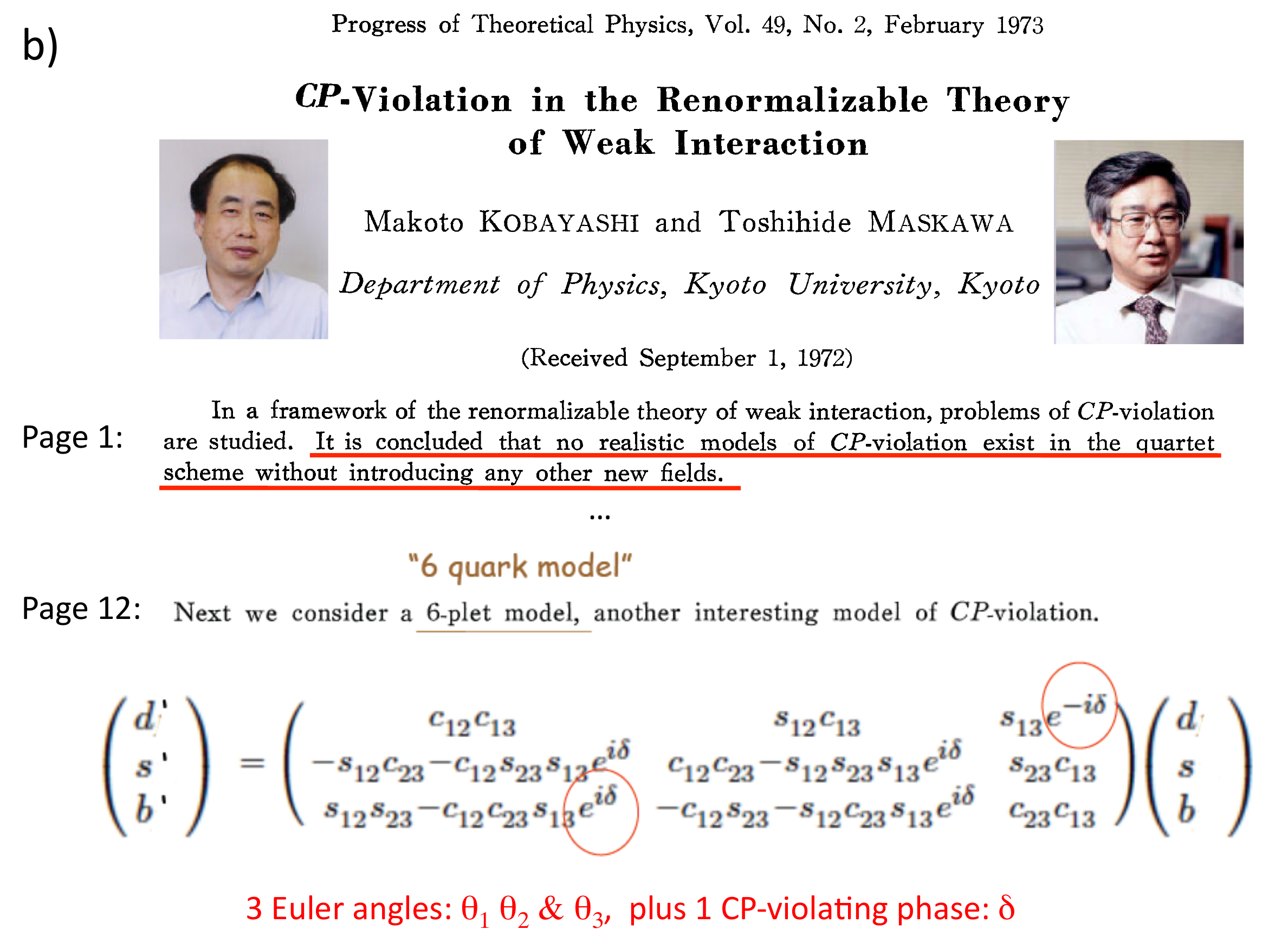}
\end{minipage}
\hspace{\fill}
\caption{\footnotesize {\bf a)}
 The amplitude ${\mathcal A}$ represents a $CP$-violating contribution
to a hypothetical process $X^0\rt\pip$. Differences in the $X^0\rt\pipi$ and $\bar{X}^0\rt\pipi$ decay rates
can only be observed if it interferes with a $CP$-conserving process (amplitude = ${\mathcal C}$) that has
a non-zero common phase $\phi_0$.  
{\bf b)}  Excerpts from page 1 (above) and page 12 (below) of the classic Kobayashi-Maskawa paper~\cite{km73}. 
}
\label{fig:cp-phase}
\end{figure}

In 1972, Kobayashi and Maskawa (KM) showed that a non-trivial $CP$-violating phase could be introduced
into the weak interaction quark-flavor mixing matrix, but only if there were at least three generations
of quark doublets, {\it i.e.}, at least six quark flavors (see Fig.~\ref{fig:cp-phase}b)~\cite{km73}.
This was remarkable because at that time, only three quark flavors had been established.
In a 1980 paper, Carter and Sanda suggested that if the $b$-quark-related flavor mixing parameters
were such that $B^0\leftrightarrow \bar{B}^0$ was substantial and the $B$-meson lifetime was relatively
long, large $CP$ violations might be observable in neutral $B$ meson decays and provide conclusive tests
of the KM idea~\cite{sanda80}.  However, the tests that Carter and Sanda proposed would require data
samples the contained several hundreds of exclusive $B^0$ decays to $CP$ eigenstates, such as
$B^0\rt K_S\jpsi$ and $B^0\rt K_S\psi'$.  In 1983, CLEO reported the world's first sample of exclusive
$B$-meson decays shown in Fig.~\ref{fig:exclusive-b}a, where there are 18 events in the $B$-meson mass peak,
divided equally between neutral and charged $B$-mesons with a background that is estimated to be between 
4~and~7~events~\cite{cleo_excl-b}.   No exclusive decays to a $CP$ eigenstate were observed.  
Thus, in the early1980's, when the state-of-the-art $\ee$ collider luminosity was $\sim 10^{31}$cm$^{-2}$s$^{-1}$,
the possibility for checking the KM idea seemed hopeless, except for a few super-optimists who could
foresee luminosities greater than $10^{33}$cm$^{-2}$s$^{-1}$ by the end of the century.

\begin{figure}[htb]
\begin{minipage}[t]{53mm}
  \includegraphics[height=0.85\textwidth,width=0.8\textwidth]{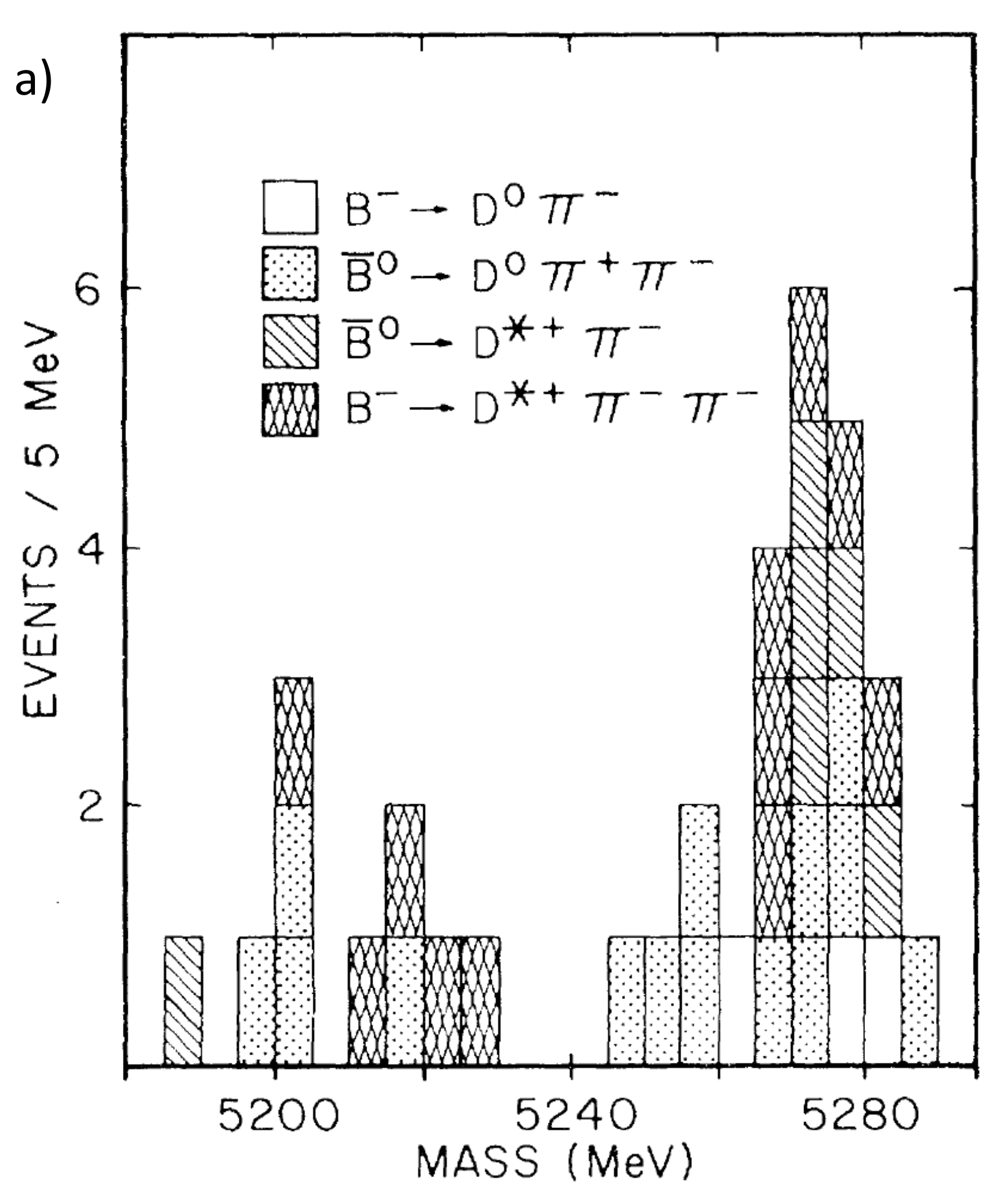}
\end{minipage}
\begin{minipage}[t]{63mm}
  \includegraphics[height=0.75\textwidth,width=1.0\textwidth]{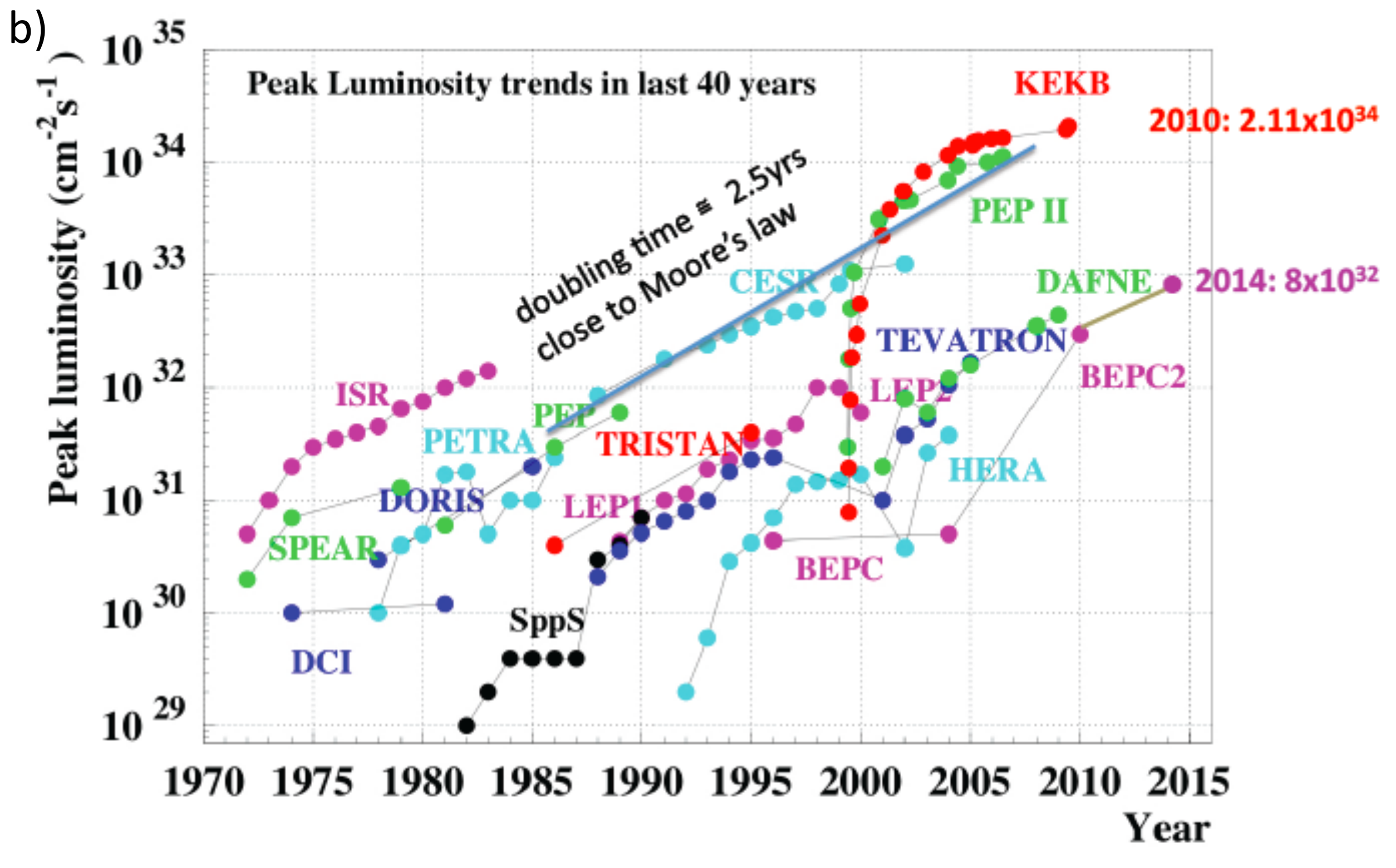}
\end{minipage}
\begin{minipage}[t]{53mm}
  \includegraphics[height=0.9\textwidth,width=0.9\textwidth]{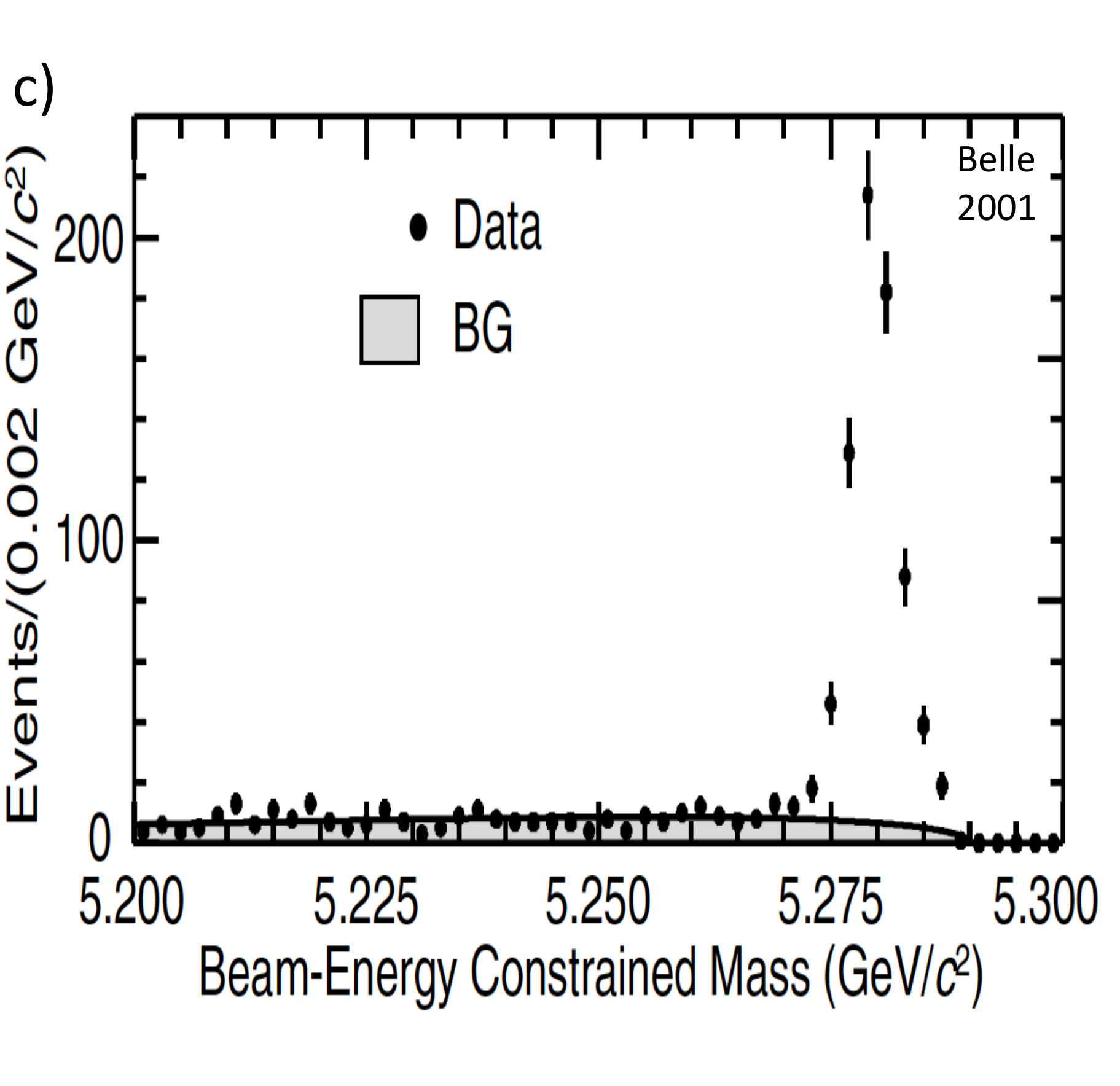}
\end{minipage}
\hspace{\fill}
\caption{
\footnotesize {\bf a)} (Figure~2 from ref.~\cite{cleo_excl-b}.) The first reported signal for exclusive
$B$-mesons decays, found by CLEO in a 40~pb$^{-1}$ data sample recorded over a three-year time
period.
{\bf b)}  A ``Livingston plot'' for $\ee$ luminosities {\it vs.} year.
{\bf c)} (Figure 1 from ref.~\cite{belle_2001}.)
The 2001 $B\rt K_S(c\cbar)$, $\xi_f=-1$ $CP$ eigenstate decay signal from a 29~fb${-1}$ Belle data sample,
containing $\sim$700~signal events, mostly $B\rt K_S J/\psi$ decays, with a 92\% signal purity.
}
\label{fig:exclusive-b}
\end{figure}

For the 1973 KM idea to be relevant and testable, there had to be: six quarks instead of three; 
a relatively long $B$-meson lifetime and sizable $B^0\leftrightarrow\bar{B}^0$ mixing (corresponding to
$|V_{cu}|<|V_{cb}|<0.1$); and a thousand-fold or more combined improvement in $\ee$ luminosity and detector
performance.  In 1974, the fourth quark, the $c$-quark, was discovered at Brookhaven~\cite{ting74} and
SLAC~\cite{richter74} and the fifth quark, the $b$-quark, was found in 1977 by a Fermilab experiment~\cite{herb77}. 
Then, in 1983, a long ($\tau_{B}\simeq 1.5$~ps) $B$-meson lifetime was measured at PEPII~\cite{mac83,pepii83} and,
in 1987, a substantial signal for $B^0\leftrightarrow\bar{B}^0$ mixing was unexpectedly discovered by the
ARGUS experiment at DESY~\cite{argus87}.  Taken together, these results indicated that the
$CKM$ mixing-angle values were favorable for experimental tests of the KM idea.  (The $B^0\leftrightarrow \bar{B}^0$
mixing frequency is now well measured to be $\omega_{0}\simeq 0.5$ps$^{-1}$, and not much different than $1/\tau_B$.)
In addition, the luminosity of $\ee$ colliders kept increasing in a Moore's-law-like fashion with a doubling time
of about 2.5~years (see Fig.~\ref{fig:exclusive-b}b).   In 2001, less than twenty years after the CLEO report
of an 18~event exclusive $B$-meson decay signal with no $CP$ eigenstate modes, the Belle experiment's discovery
paper on $CP$-violation in the $B$-meson system used the $\sim 700$ neutral $B$ mesons to $CP$ eigenstate decays
with $CP$ eigenvalue $\xi_f =-1$ (mostly $B\rt K_S J/\psi$) in the signal peak shown in
Fig.~\ref{fig:exclusive-b}c~\cite{belle_2001}. 

\begin{figure}[htb]
\begin{minipage}[t]{87mm}
  \includegraphics[height=0.5\textwidth,width=0.8\textwidth]{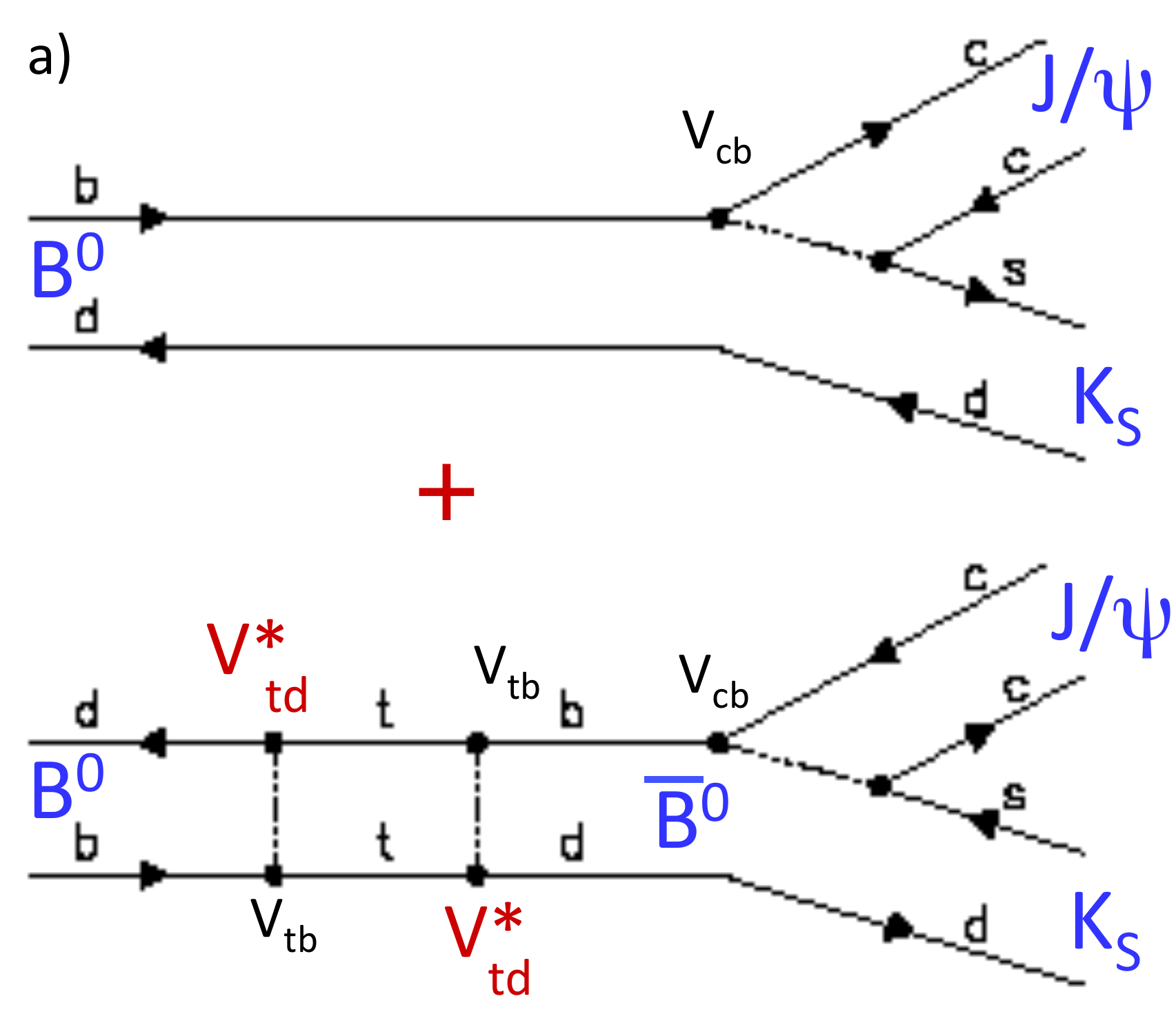}
\end{minipage}
\begin{minipage}[t]{87mm}
  \includegraphics[height=0.5\textwidth,width=0.8\textwidth]{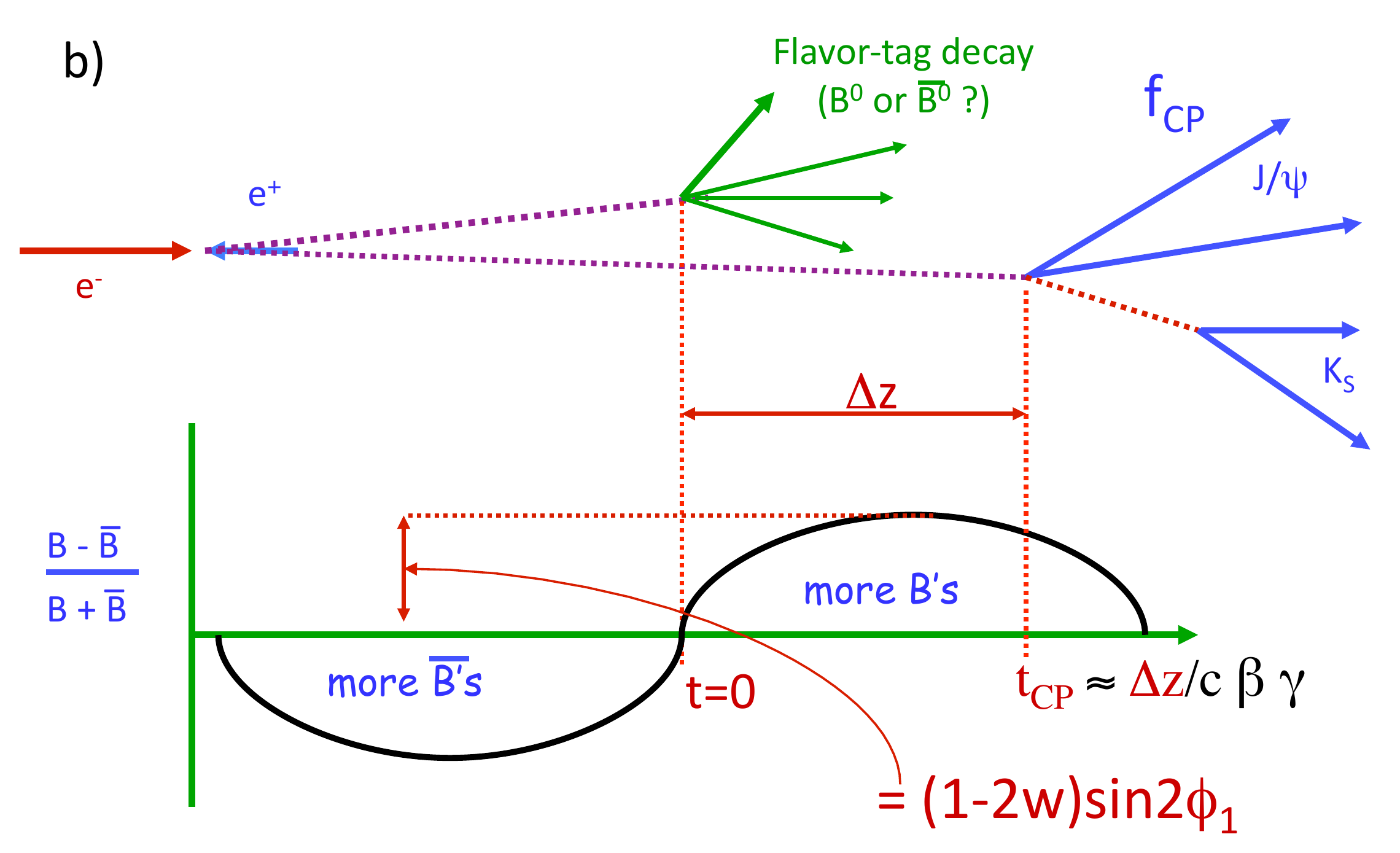}
\end{minipage}
\hspace{\fill}
\caption{ \footnotesize {\bf a)} A $B^0$-meson can decay to a $CP$ eigenstate directly or by first mixing into a $\bar{B}^0$
that in turn decays the same $CP$ eigenstate.  The interference between the two processes is $\propto V_{td}^{*2}$
(not $|V_{td}^*|^2$). 
{\bf b)}  A cartoon that illustrates how the $B$-factory experiments measure $\phi_1$, the $CP$-violating phase of $V_{td}^*$.
}
\label{fig:cp-meas}
\end{figure}

Carter and Sanda suggested that $\phi_1$, the $CPV$ phase of $V_{td}$, could be measured by the interference
between the two $B^0$-meson quark-line processes shown in Fig.~\ref{fig:cp-meas}a.  Here the top diagram is the direct
$B^0$-meson decay to a $CP$ eigenstate (chosen here as $K_S J/\psi$ for illustration). In the lower diagram, the $B^0$
first mixes into a $\bar{B}^0$ and the $\bar{B}^0$ decays to the same $CP$ eigenstate.  The amplitude for
the upper diagram is proportional to $V_{cb}$, which has no $CPV$ phase; that for the lower diagram is
proportional to $V_{tb}^2V_{td}^{*2}V_{ub}$, where, in the KM formalism, only $V_{td}^{*}$ has a $CPV$ violating phase.
Thus, the interference term is $\propto V_{td}^{*2}\propto\sin 2\phi_1$.

The way this interference is measured is illustrated in Fig.~\ref{fig:cp-meas}b.  An asymmetric energy $\ee$ collison 
produces a boosted $B^0$ and a $\bar{B}^0$ in a ``entangled'' $J^{PC}=1^{--}$ quantum state.  After some time, 
one of the $B$ meson decays to a ``flavor specific'' final-state, {\it i.e.} a final state that allows one to distinguish
whether the flavor of decaying $B$ meson is a $B^0$ or a $\bar{B}^0$.  At that time, which is taken as $t=0$, the accompanying
$B$ meson has to have the opposite flavor.  Then this accompanying $B$ meson evolves with time, mixing as it goes along
(either forward or backward in time!) into
the opposite flavor with a frequency $\omega_{\rm mix}$, and eventually decays at time $t$ into a $CP$ eigenstate. What is measured,
is the asymmetry ${\mathcal A}_{CP}$ as a function of $t$, where 
\begin{equation}
{\mathcal A}_{CP}=\frac{N_{\bar{B}^0}-N_{B^0}}{N_{\bar{B}^0}+N_{B^0}}=\xi_{f}(1-2w)\sin 2\phi_1\sin\omega_B t,
\end{equation}
$N_{B^0}$ ($N_{\bar{B}^0}$) is the number of times the flavor-tagged $B$ is a $B^0$ ($\bar{B}^0$), $\xi_{f}$ is the $CP$
eigenvalue of the state being studied (for $B\rightarrow K_S J/\psi$, $\xi_{f}=-1$), $w$ is the probability that the
flavor-tagged $B$ meson is assigned the wrong flavor, and $t$ is inferred from $\Delta z$, the measured separation
of the two $B$-meson decay vertices: $t=\Delta z/(c\gamma\beta)$.  In this measurement, the required common phase that was
discussed above in conjunction with Fig.~\ref{fig:cp-phase}a is provided by the mixing term $\exp (i\omega_B t)$, which changes
sign at $t=0$.  Thus the time integrated asymmetry is zero and the boost ($\gamma\beta$) provided by the energy asymmetry of
the beams is essential.

At the time the flavored-tagged $B$-meson decays, the accompanying $B$ meson is in a pure flavor state,
and the interference (and asymmetry) is zero; as this meson propagates, its flavor mixes and, after
about 3~ps, the $B^0$ and $\bar{B}^0$ amplitudes are nearly equal and the asymmetry is maximum.  However,
this 50:50 mixing occurs for a decay-time difference of about two $B^0$ meson lifetimes, and only $\sim$13\% of
the $B$ mesons live this long.  This, and the small branching fractions for $B^0$ decays to measureable
$CP$ eigenstates (typically $\sim 0.1$\%), explain why such a huge increase in $\ee$ collider luminosity was critical
for these measurements.

The $K_L J/\psi$ final state has $\xi_{f}=+1$ and a $CPV$ asymmetry that is opposite in sign to that for
$K_S J/\psi$ final states.  Thus, both BaBar and Belle instrumented their magnet return yoke to make it suitable for
reconstructing $K_L J/\psi$ final states.  A $K_L$ can produce a splash of energy in the instrumented return yoke,
either by decaying or interacting in one of the yoke's iron plates, as shown
in the top panel of Fig.~\ref{fig:cp-results}a, that can be used to determine the $K_L$ direction.  That, with the
assumption of two-body decay dynamics, can be used to infer $p_{B}^{\rm cms}$, the $B$ meson's three-momentum
in the center of mass (c.m.) system.  Belle's 2001 $p_{B}^{\rm cms}$ distribution, shown in the lower panel of
Fig.~\ref{fig:cp-results}a, exhibits a distinct, $\sim$346-event signal peak for $B\rightarrow K_L J/\psi$ decays at
$p_{B}^{\rm cms}\simeq 0.33$~GeV/$c$ (with a 61\% signal purity) that were also used for $CPV$ asymmetry measurements.  
\begin{figure}[htb]
\begin{minipage}[t]{55mm}
  \includegraphics[height=0.9\textwidth,width=0.8\textwidth]{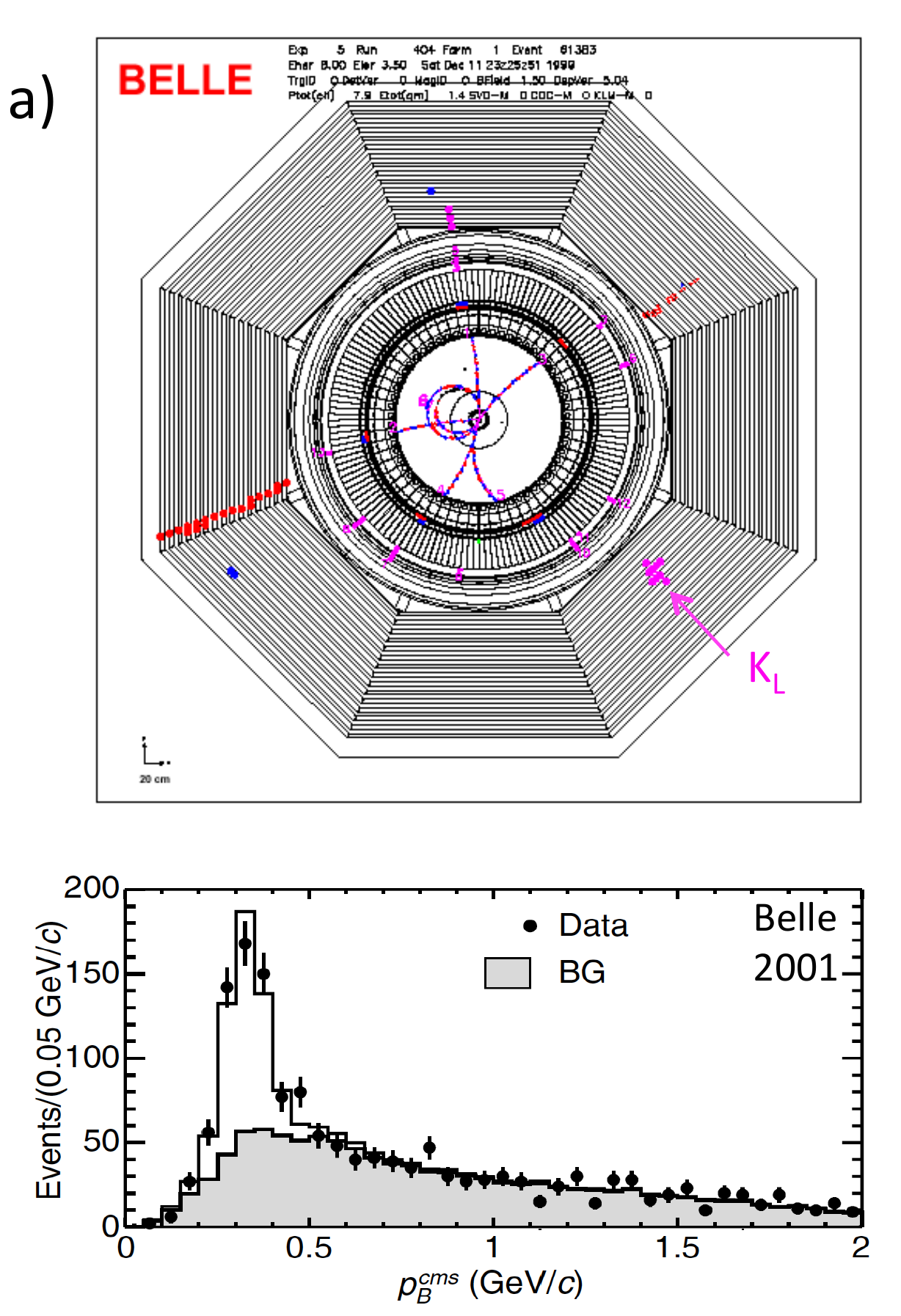}
\end{minipage}
\begin{minipage}[t]{55mm}
  \includegraphics[height=0.9\textwidth,width=0.8\textwidth]{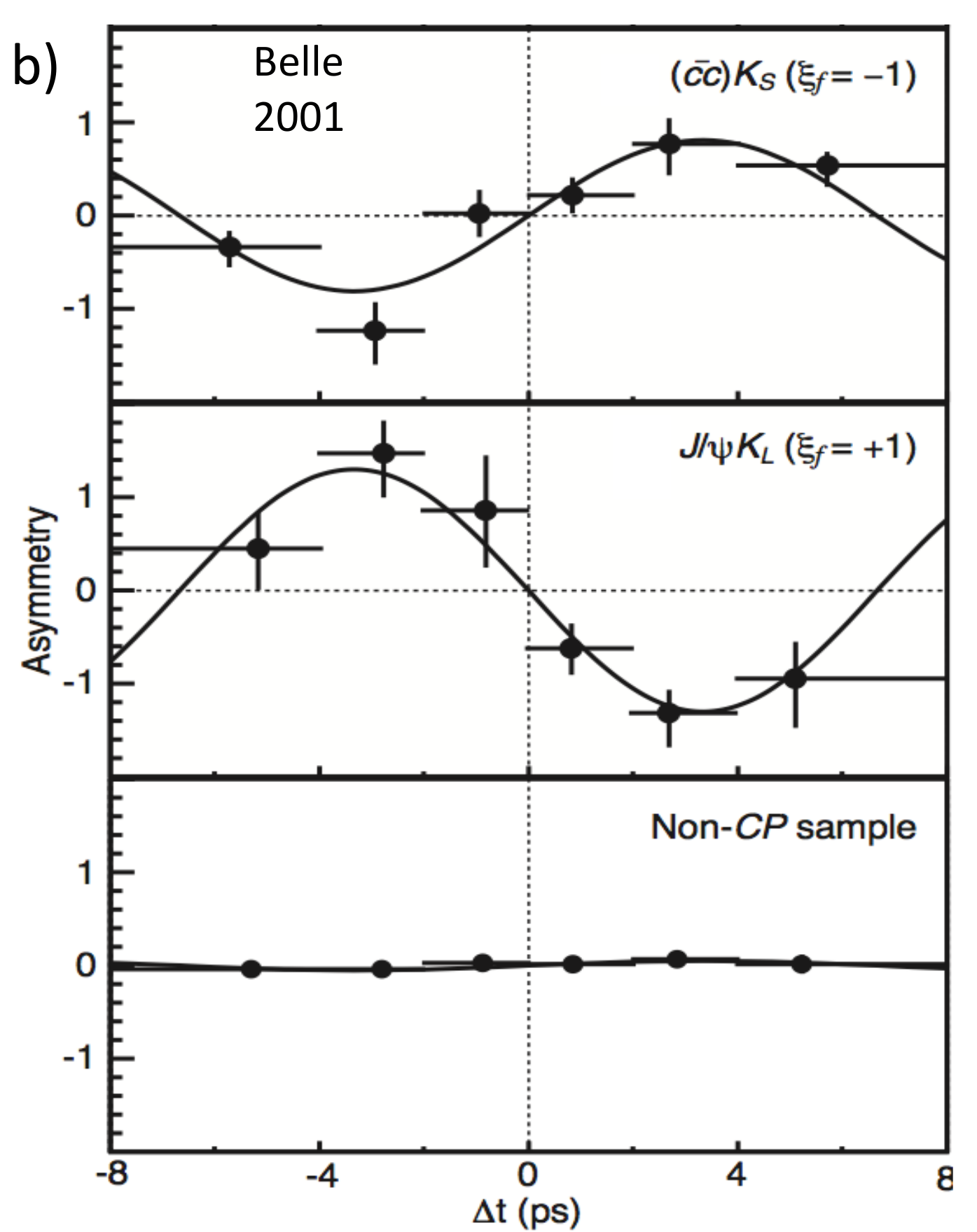}
\end{minipage}
\begin{minipage}[t]{55mm}
  \includegraphics[height=0.9\textwidth,width=0.8\textwidth]{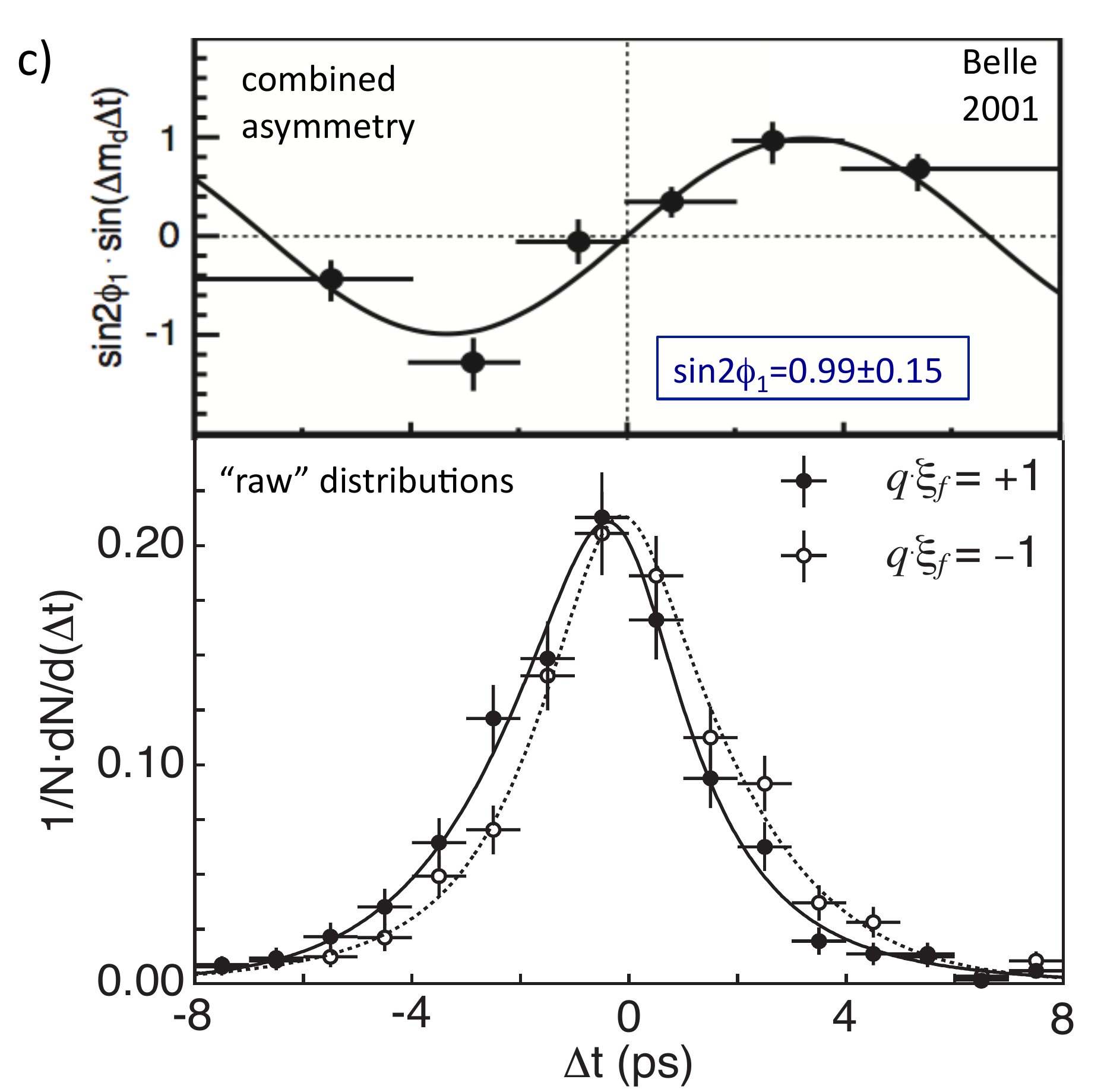}
\end{minipage}
\hspace{\fill}
\caption{ \footnotesize {\bf a)} (top) A computer display of $B\rightarrow K_L J/\psi$ events candidate in Belle,
where the $J/\psi$ decays into a $\mumu$ pair and the $K_L$ produces an energy cluster in the magnet's instrumented
flux return.  (bottom) The $p_{B}^{\rm cms}$ distribution for candidate $K_L J/\psi$ events.
{\bf b)} The $t$-dependent $CPV$ asymmetry for $\xi_f =-1$ (top), $\xi_f =+1$ (center) and non-$CP$ eigenstate
decays (bottom).   {\bf c)} The $t$ dependence of events organized according to $q\xi_f$ values, where $q=+1$ ($-1$)
corresponds to a tagged $B^0$ ($\bar{B}^0$) (bottom).  The combined $q\xi_f =-1$ minus $q\xi_f=+1$ asymmetries, together
with the fit results (top).  
}
\label{fig:cp-results}
\end{figure}

The 2001 Belle result, $\sin 2\phi_1 =0.99\pm 0.15$~\cite{belle_2001}, was 6$\sigma$ from zero and conclusively confirmed
the KM prediction for a non-zero $CPV$ complex phase in the $V_{td}$ element of the quark-flavor mixing matrix. The
opposite asymmetries for $\xi_{f}=+1$~and~$-1$ decay samples, shown in the center and top panels of Fig.~\ref{fig:cp-results}b, 
respectively, provided a check on possible systematic effects on the $\sin 2\phi_1$ measurements.
Another validity check is illustrated in the lower panel of Fig.~\ref{fig:cp-results}b, which shows the results of the
same analysis applied to non-$CP$ eigenstate decay modes, where no asymmetry is expected; the fit result for these
events is $0.05\pm 0.04$.  At the same time, the BaBar experiment reported a 4$\sigma$ non-zero value:
$\sin 2\phi_1 = 0.59\pm 0.15$~\cite{babar_2001}.

The combined average of the 2001 BaBar and Belle $\phi_1$ results is compared with constraints from
other measurements in Fig.~\ref{fig:ckm-fit}a ~\cite{ckm-fitter}, where good agreement with
expectations is evident. Eventually BaBar and Belle each accumulated a huge amount 
of additional data and significantly improved the precision on their $\phi_1$ measurements and
other quantities that now constrain the 2015 allowed region of the same plane~\cite{babar-belle}
as shown in Fig.~\ref{fig:ckm-fit}b, which demonstrates that the consistency of the CKM picture is
amazingly good.  This success resulted in Kobayahi and Maskawa sharing the 2008 Physics Nobel prize
(with Yoichiro Nambu).

\begin{figure}[htb]
\begin{minipage}[t]{55mm}
  \includegraphics[height=0.8\textwidth,width=0.9\textwidth]{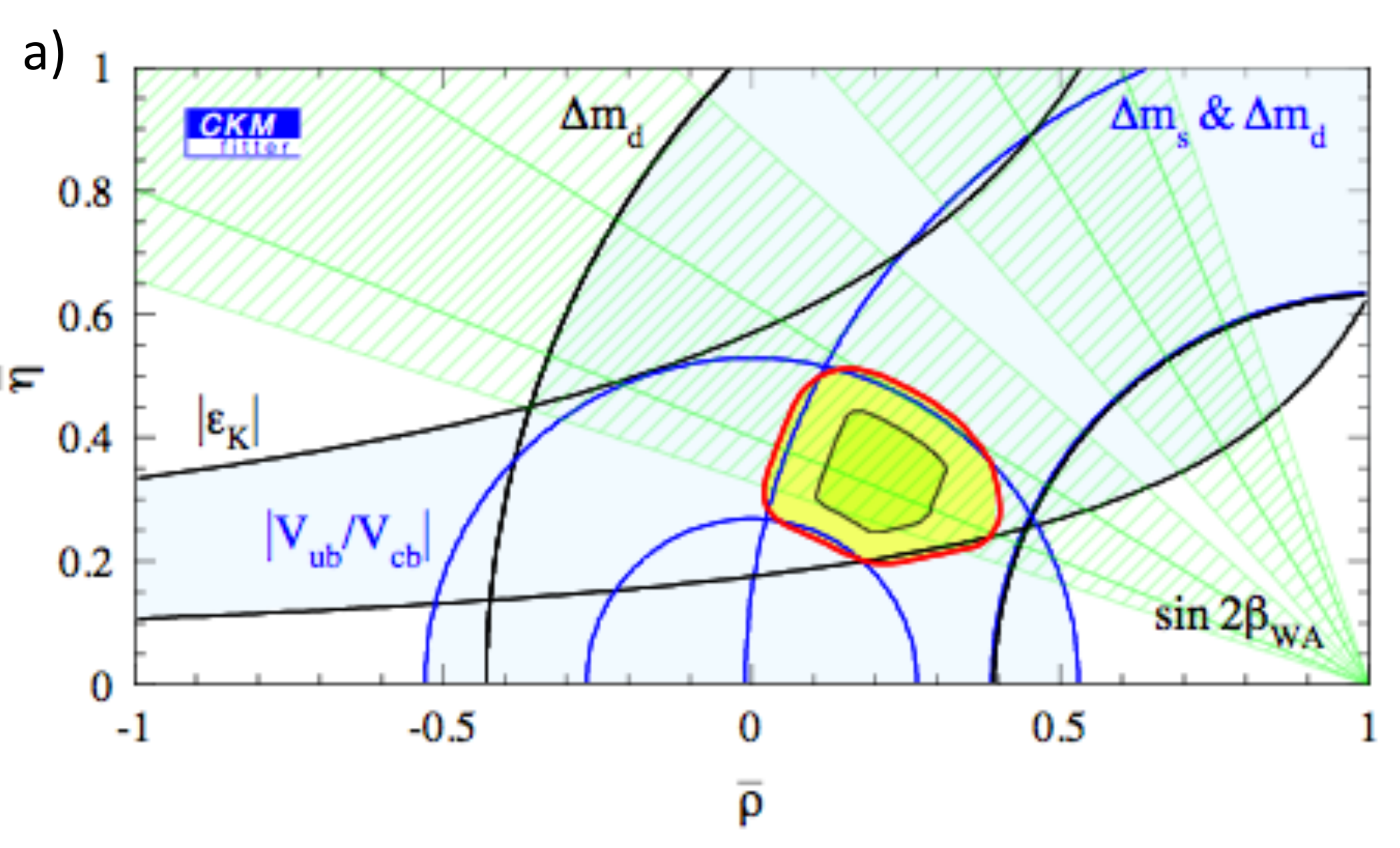}
\end{minipage}
\begin{minipage}[t]{55mm}
  \includegraphics[height=0.8\textwidth,width=0.9\textwidth]{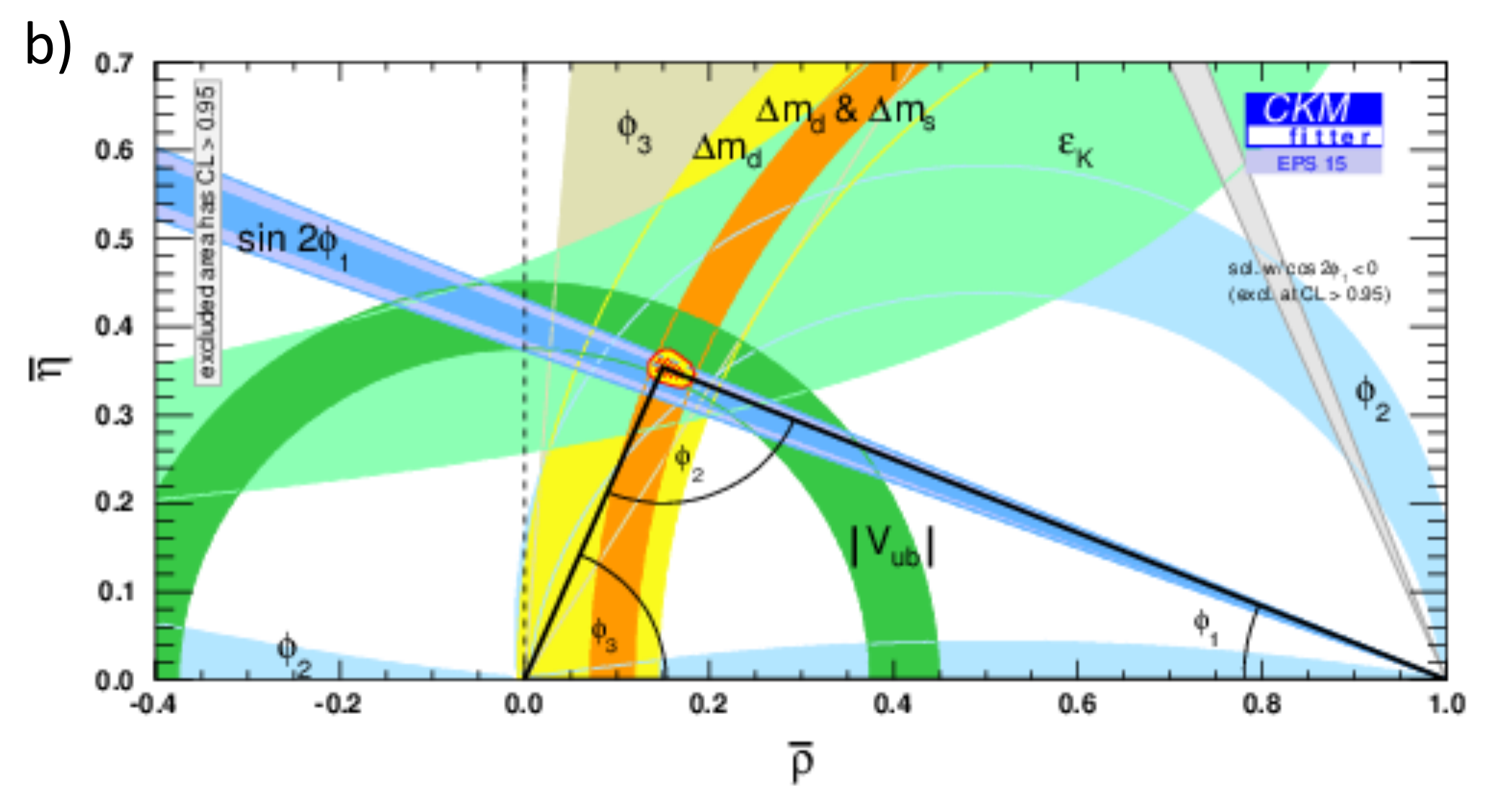}
\end{minipage}
\begin{minipage}[t]{55mm}
  \includegraphics[height=0.8\textwidth,width=0.9\textwidth]{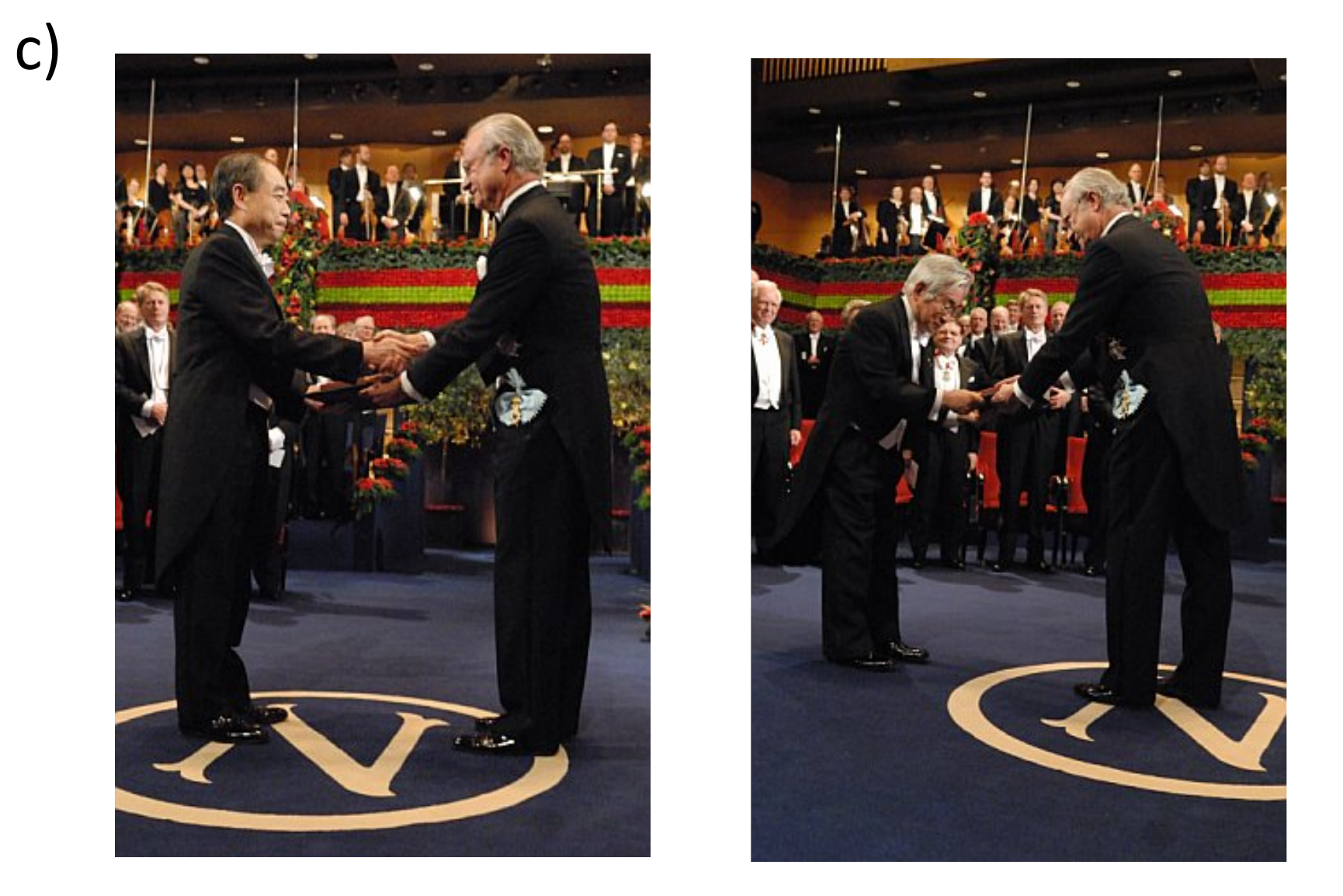}
\end{minipage}
\hspace{\fill}
\caption{ \footnotesize {\bf a)}  The unitarity triangle plot from the CKM-fitter group with the average of
the 2001 BaBar and Belle $\sin 2\phi_1$ results (labeled as $\sin 2\beta_{WA}$). Here $\bar{\rho}$ and $\bar{\eta}$  
are the Wolfenstein $CPV$ parameters~\cite{wolfenstein83}. {\bf b)}  The 2015 version of the CKM-fitter group's unitary triangle plot.
{\bf c)}  (left) Kobayashi and (right) Maskawa meeting the King of Sweden in Dec. 2008.
}
\label{fig:ckm-fit}
\end{figure}

\section{It wasn't only about $\mathbf{CP}$, or even $\mathbf B$ mesons}

\subsection{Double $\mathbf{c\bar{c}}$ production in $\mathbf{\ee}$ annihilation }
One of the earliest measurements in Belle was a study of inclusive $J/\psi$ production in 
continuum $\ee$ annihilation at c.m. energies near 10.6~GeV.  Studies of $J/\psi$ production
is a common activity for the early stages of an experiment because they are a prolific source of
tagged muons and electrons that are useful for calibrating lepton identification systems, validating
triggers and tuning up charged particle tracking algorithms.  Theoretically, inclusive
and exclusive $J/\psi$ production is supposed to be described accurately (and rigorously) by
non-relativistic quantum chromodynamics, NRQCD~\cite{bodwin95}.

In 2002, Belle reported a total cross section for the inclusive, continuum annihilation process
$\ee\rightarrow J/\psi +X$ of $1.47\pm 0.16$~pb~\cite{belle_4c2002}. This was in reasonable
agreement with NRQCD~\cite{schuler99}, which had predicted a $\sim$1.1~pb cross section that is
$\sim$(1/3)$^{\rm rd}$ due to $\ee\rightarrow gg (c\bar{c})_1 $ and $\sim$(2/3)$^{\rm rds}$ due
to $\ee \rightarrow g (c\bar{c})_8$, where $(c\bar{c})_1$ and $(c\bar{c})_8$ refer to color-singlet
and color-octet charmed-quark anticharmed-quark configurations, respectively.

However, Belle's measured $J/\psi$ momentum distribution, shown in Fig.~\ref{fig:cccc}a, has
no significant event signal in the highest kinematically allowed momentum region, $4.5$-$4.84$~GeV/$c$,
where the dominant color-octet contribution was expected to be strongest.  A MC estimate for the number 
of expected signal events in this high momentum region using a special NRQCD-inspired event generator
incorporated into PYTHIA~\cite{pythia} predicted a $\sim$300 event signal in the two highest bins of
Fig.~\ref{fig:cccc}a, where no signal is seen.

\begin{figure}[htb]
\begin{minipage}[t]{55mm}
  \includegraphics[height=0.8\textwidth,width=0.9\textwidth]{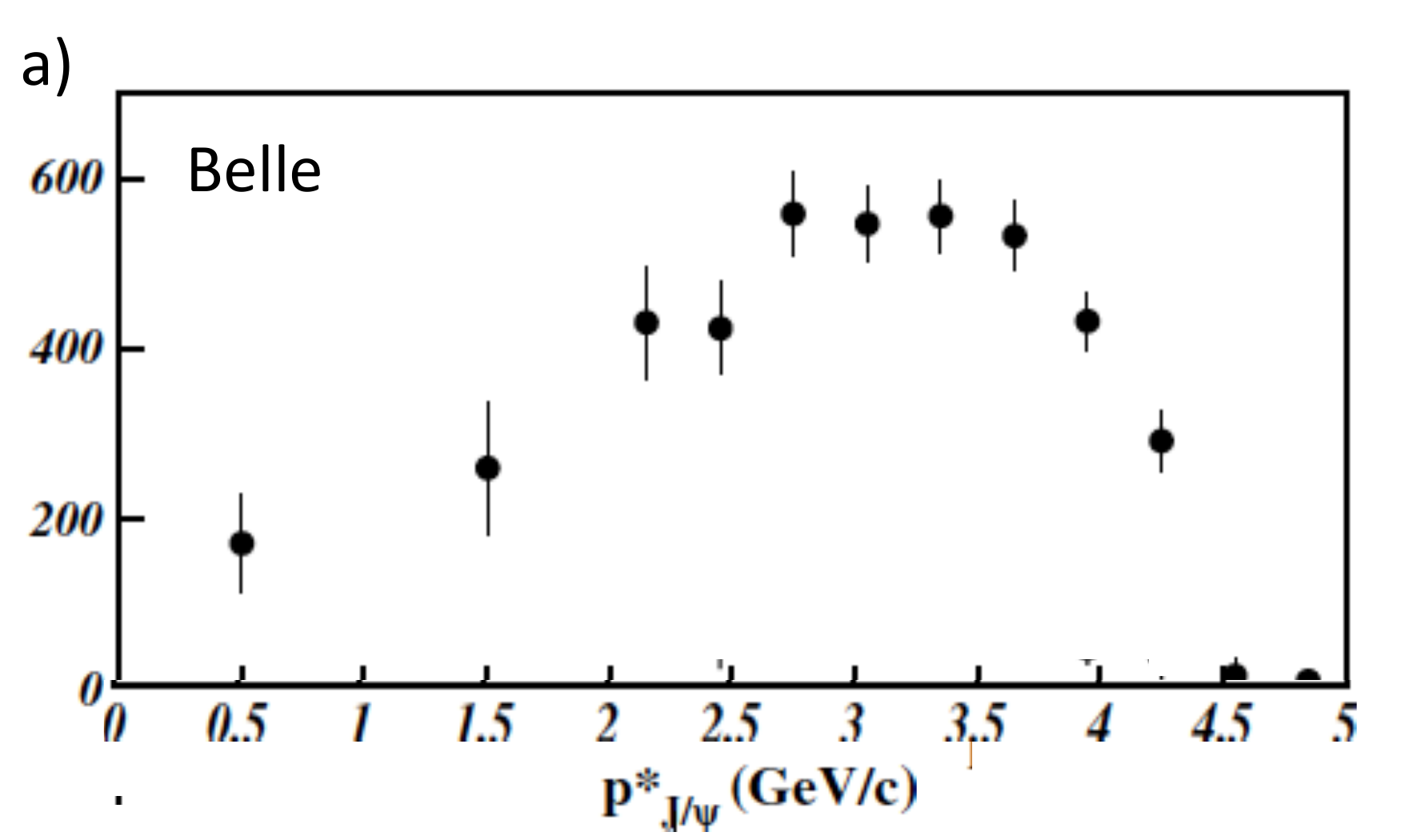}
\end{minipage}
\begin{minipage}[t]{55mm}
  \includegraphics[height=0.8\textwidth,width=0.9\textwidth]{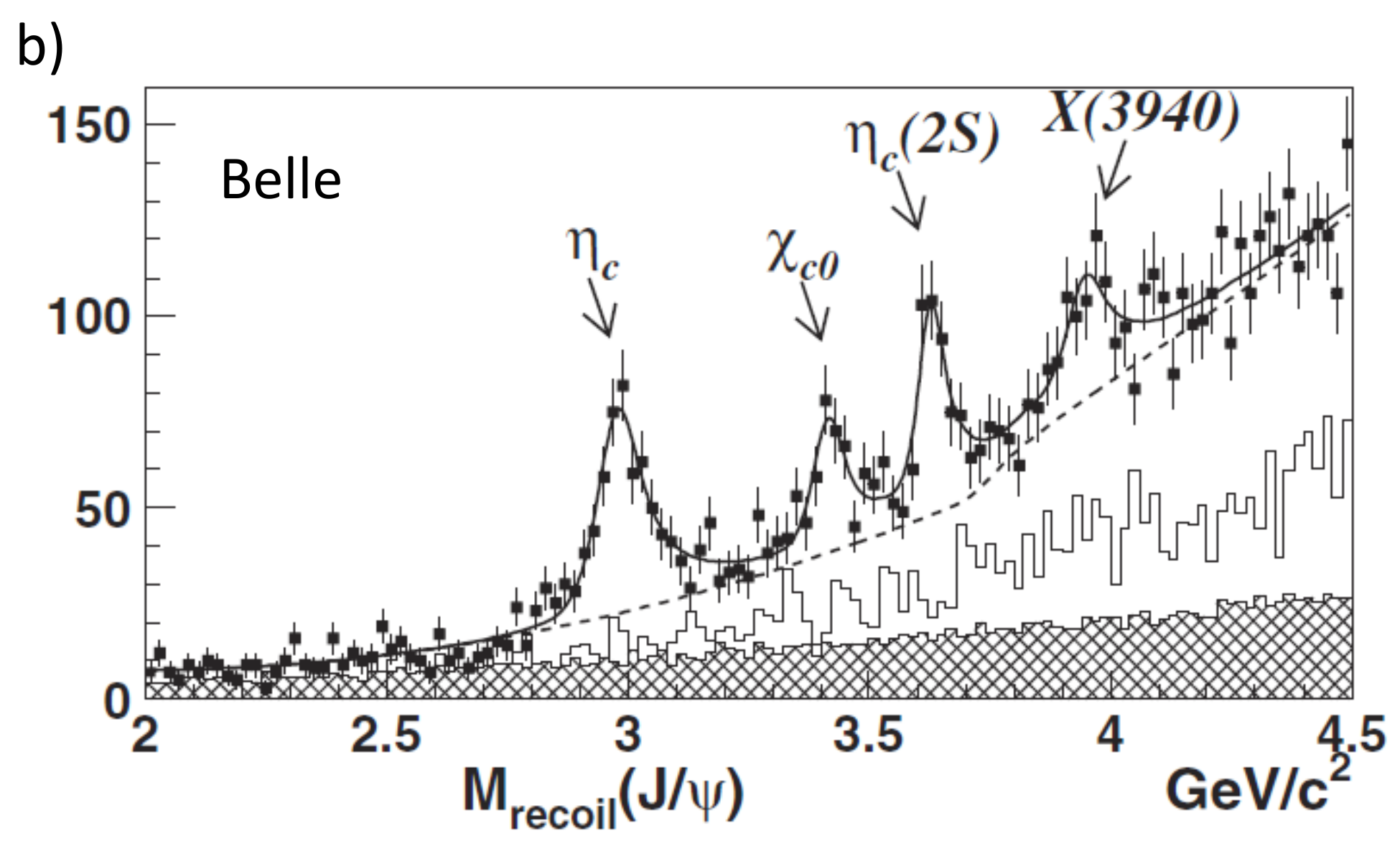}
\end{minipage}
\begin{minipage}[t]{55mm}
  \includegraphics[height=0.8\textwidth,width=0.9\textwidth]{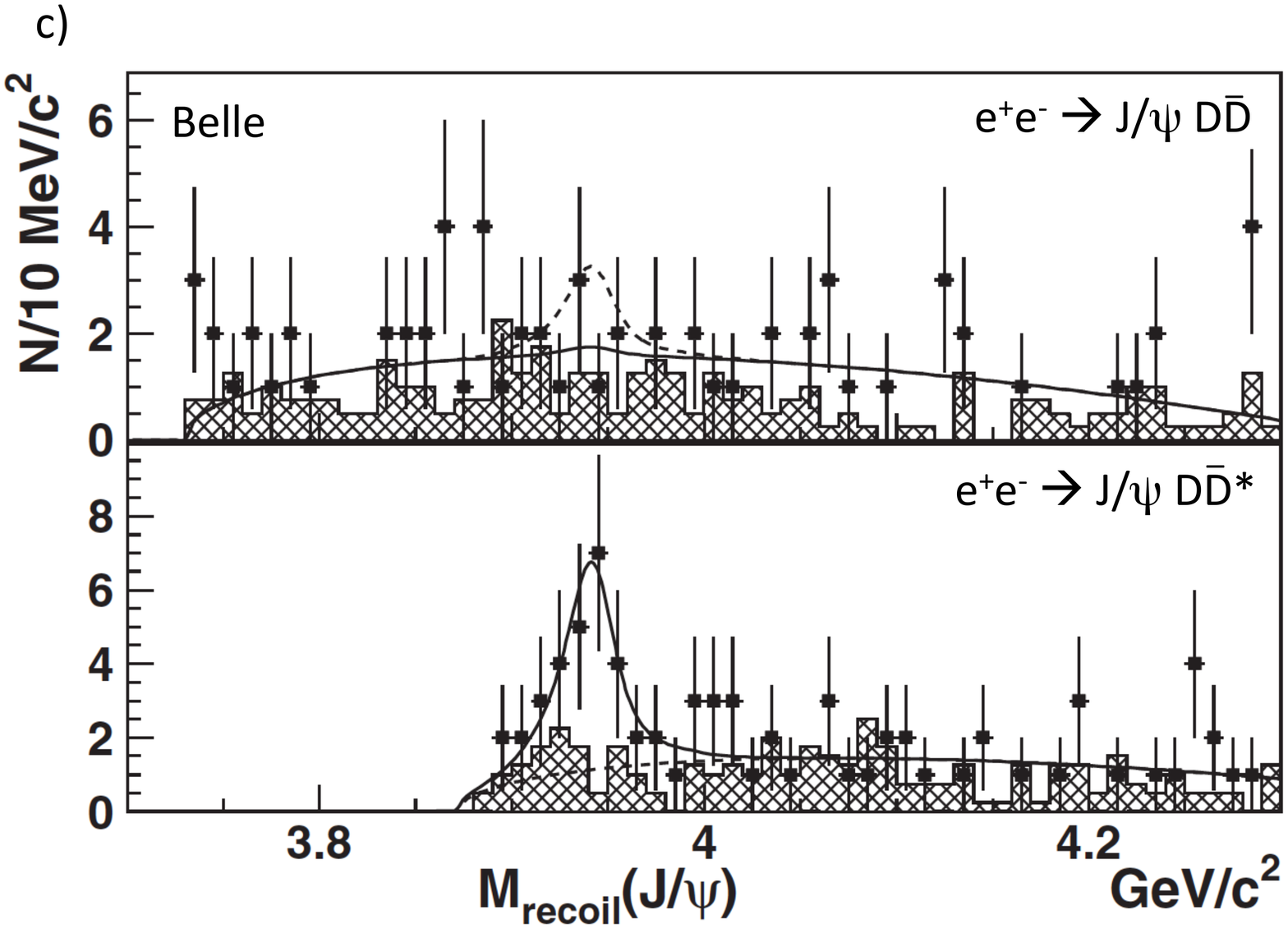}
\end{minipage}
\hspace{\fill}
\caption{ \footnotesize {\bf a)}  The $J/\psi$ c.m. three-momentum distribution
for inclusive $\ee\rightarrow J/\psi X$  reactions near $\sqrt{s}= 10.6$~GeV
(from ref.~\cite{belle_4c2002}).
{\bf b)}  The distribution of masses recoiling from the $J/\psi$ in inclusive 
$\ee\rightarrow J/\psi X$ annihilations. The shaded
histogram is background estimated from the $J/\psi$ mass side bands; the open histogram
is the feed down from $\psi ^{\prime}\rightarrow J/\psi + X$ (from ref.~\cite{belle_4c2007}).
{\bf c)} The $J/\psi$ recoil mass distributions for $\ee\rightarrow J/\psi D\bar{D}$
(upper) and $\ee\rightarrow J/\psi D\bar{D}^*$ (lower) events (from ref.~\cite{belle_4c2007}).
The hatched histogram shows the background estimated from the $D$-mass sidebands. (The inclusion
of charge-conjugate states is implied.)}
\label{fig:cccc}
\end{figure}

A 2007 Belle study of the same process with more data, reported results in terms of the mass recoiling from
the detected $J/\psi$ ({\it i.e.} $M_{\rm recoil}(J/\psi) = \sqrt{(E_{cms}-E^{cms}_{J/\psi})^2-p^{cms}_{J/\psi}}~~)$ shown
in Fig.~\ref{fig:cccc}b~\cite{belle_4c2007}.  This distribution has a number noteworthy features:
\begin{itemize}
\item  there are no obvious signal events below the $\eta_c$ peak, where contributions from color-octet production are
       expected to be strongest; 
\item the $\sim$500 event $\eta_c$ signal corresponds to a cross section for the exclusive process
       $\ee\rightarrow J/\psi \eta_c$ of $25.6\pm 4.4$~fb~\cite{belle_4c2004}, more than an order
       of magnitude higher than NRQCD-based expectations~\cite{braaten03,chao03};
\item the $\sim 300$~event $\eta_c(2S)$ signal provided the best confirmation of this state at that time;
\item the three lower-mass peaks all correspond to established, spin=0 charmonium states;
\item there is strong production ($\sigma\simeq 10$~fb) of a previously unknown state with $M\simeq 3940$~MeV.
\end{itemize} 

Belle found that the $J/\psi c\bar{c}$ component corresponds to $(59\pm 0.18)$\% of the total inclusive
$J/\psi$ production cross section~\cite{belle_4c2002a} in contradiction to NRQCD expectations that
it would be $\lesssim$10\% of the $J/\psi gg$ component~\cite{kiselev94,berezhnoy04}. The
cross sections  for exclusive double-charmonium processes (such as $J/\psi \eta_c$) are well above
lowest-order NRQCD-based predictions~\cite{braaten03,chao03}.  This inspired studies of the corrections
due the next-to-leading order (NLO) ~\cite{chao06,gong08,bodwin08}, and these were found to be large (large enough
to explain the discrepancy), but such large corrections at NLO raise suspicions about the convergence of the NRQCD
expansion~\cite{bodwin14}.   

Belle's experimental results on double $c\bar{c}$ production have had (and are still having) a huge impact on the
development of NRQCD and, although they are not well known outside of this specialty, they are very important to,
and highly cited by, practioners in this field. (At the end of 2015, refs.~\cite{belle_4c2007},~\cite{belle_4c2004}
and~\cite{belle_4c2002a} had 271,~155~and~320 citations, respectively.)

\subsubsection{The mass peak at 3940~MeV}
In order to study the peak at 3940~MeV in the $J/\psi$ recoil mass spectrum, Belle selected events with a reconstructed
$J/\psi$ and $D$ meson~\cite{belle_4c2007}.  In these events the distribution of masses recoiling from the
$J/\psi$-$D$ system exhibit clear and distinct signals for recoil $\bar{D}$ and $\bar{D}^*$ mesons.  The $D\bar{D}$
and $D\bar{D}^*$ invariant mass distributions  for these events are shown in the upper and lower panels,
respectively, of Fig.~\ref{fig:cccc}c,  where a clear peak at $3.94$~GeV is evident in the $D\bar{D}^*$
spectrum but not in the $D\bar{D}$ channel.  
 
The absence of any signals for known spin=1 or spin=2 charmonium states in the $J/\psi$ recoil mass
spectrum of Fig.~\ref{fig:cccc}b, and the lack of any significant signal for the the $3940$~MeV peak
in the $D\bar{D}$ mass distribution in Fig.~\ref{fig:cccc}C (upper), provide circumstantial evidence
that the $J^{PC}$ quantum numbers for this new state are $0^{-+}$, which would make it a candidate
for the $\eta_c(3S)$ charmonium state.  However, in this case its $3942\pm 9$~MeV mass would be
$\sim$100~MeV below its hyperfine partner, the $\psi(3S)=\psi(4040)$, implying a hyperfine
splitting that is about twice as large as the $\psi(2S)$-$\eta_c(2S)$ splitting.  This is
contrary to expectations from potential models in which the hyperfine splitting decreases with
increasing radial quantum number. For states above open charmed thresholds, na\"{i}ve potential model
results are modified by the influence of coupled pairs of open-charmed mesons. The nearest
open charmed pair relevant to the $\eta_c(3S)$-$\psi(3S)$ doublet is a $D\bar{D}^*$ system in a
relative $P$-wave, and this should not have a very large effect on the hyperfine splitting, which is primarily
sensitive to the $c\bar{c}$ wave function at the origin. These issues are discussed in ref.~\cite{elq06}.

\subsection{Probing the $\mathbf{f_0(980)}$ and $\mathbf{a_0(980)}$scalar mesons}

The nature of the scalar mesons with mass below 1~GeV is one of the most long-standing mysteries
of hadron physics.  Although they have been studied for more than four decades, they continue to
remain controversial~\cite{amsler04,achasov08}.  It has been suggested that they are not ``standard''
$q\bar{q}$ mesons but, instead, four quark states either of the diquark-diantiquark~\cite{jaffe77},
or meson-meson molecule~\cite{voloshin76,isgur83,tornqvist92} variety. 

A way to distinguish between different substructures proposed for the scalar mesons
is by the determinations of the two-photon widths ($\Gamma_{\gamma\gamma}$) of the electrically neutral
$f_0(980)$ and $a_0^0(980)$ states via measurements of their production cross sections in $\gamma\gamma$
collisions.  Figure~\ref{fig:f0980}a illustrates how this works for $q\bar{q}$ mesons.
Both photons couple to the internal quark pair and the  partial-widths are proportional
the $e_q^4$.  Thus, for example, in the $q\bar{q}$ picture for the isoscalar $f_0(980)$ meson, (where
$q=u~{\rm and}~d$), the expectation for $\Gamma_{\gamma\gamma}(f_0(980))$ is in the range
1.3~to~1.8~keV~\cite{munz96}; for a four-quark $K\bar{K}$ molecule it is more complicated and much smaller,
in the 0.2-0.6~keV range~\cite{barnes92}; for $s\bar{s}$ it is expected to be in the range 0.3-0.5~keV~\cite{oller98}.

\begin{figure}[htb]
\begin{minipage}[t]{55mm}
  \includegraphics[height=0.8\textwidth,width=0.9\textwidth]{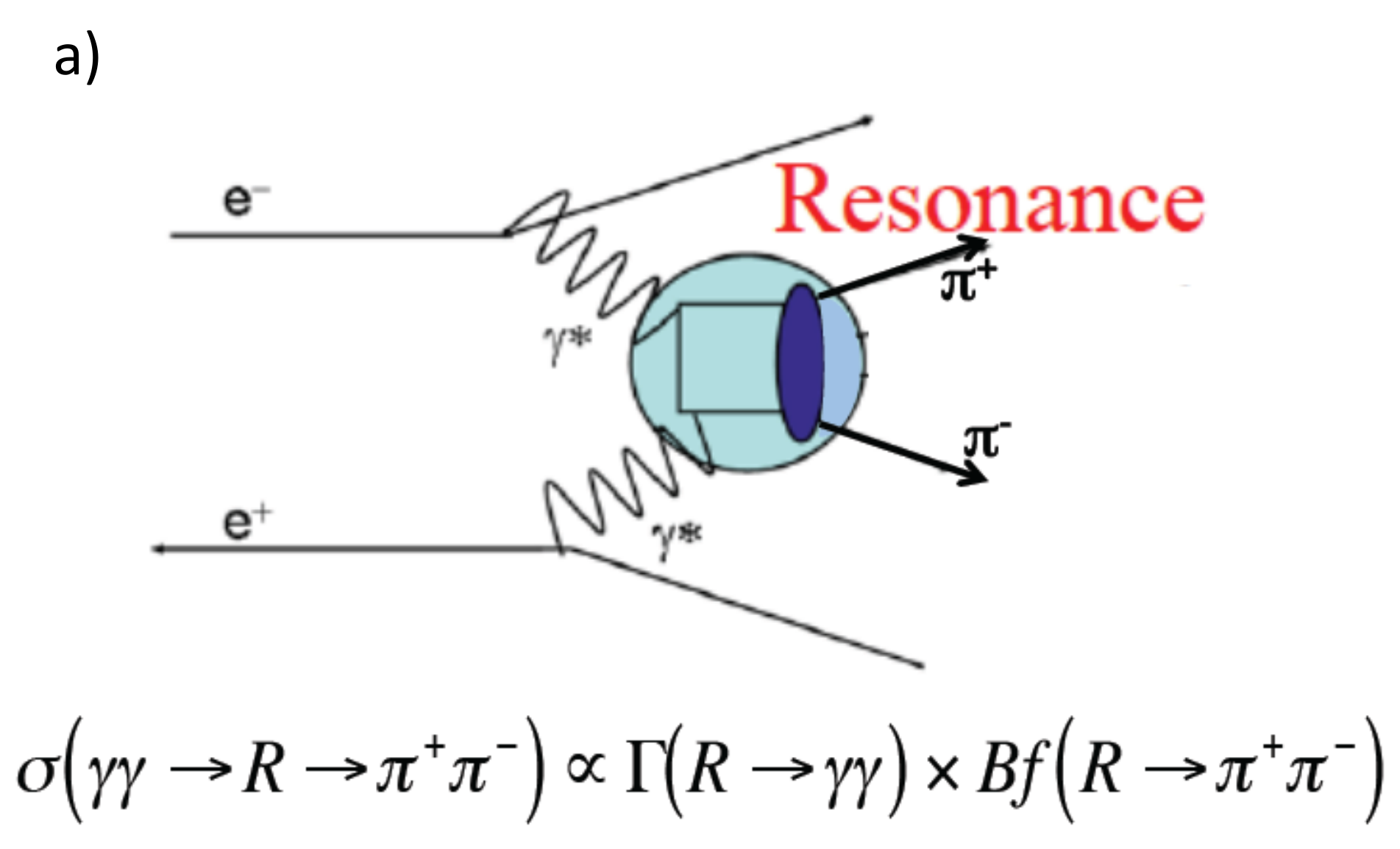}
\end{minipage}
\begin{minipage}[t]{55mm}
  \includegraphics[height=0.8\textwidth,width=0.9\textwidth]{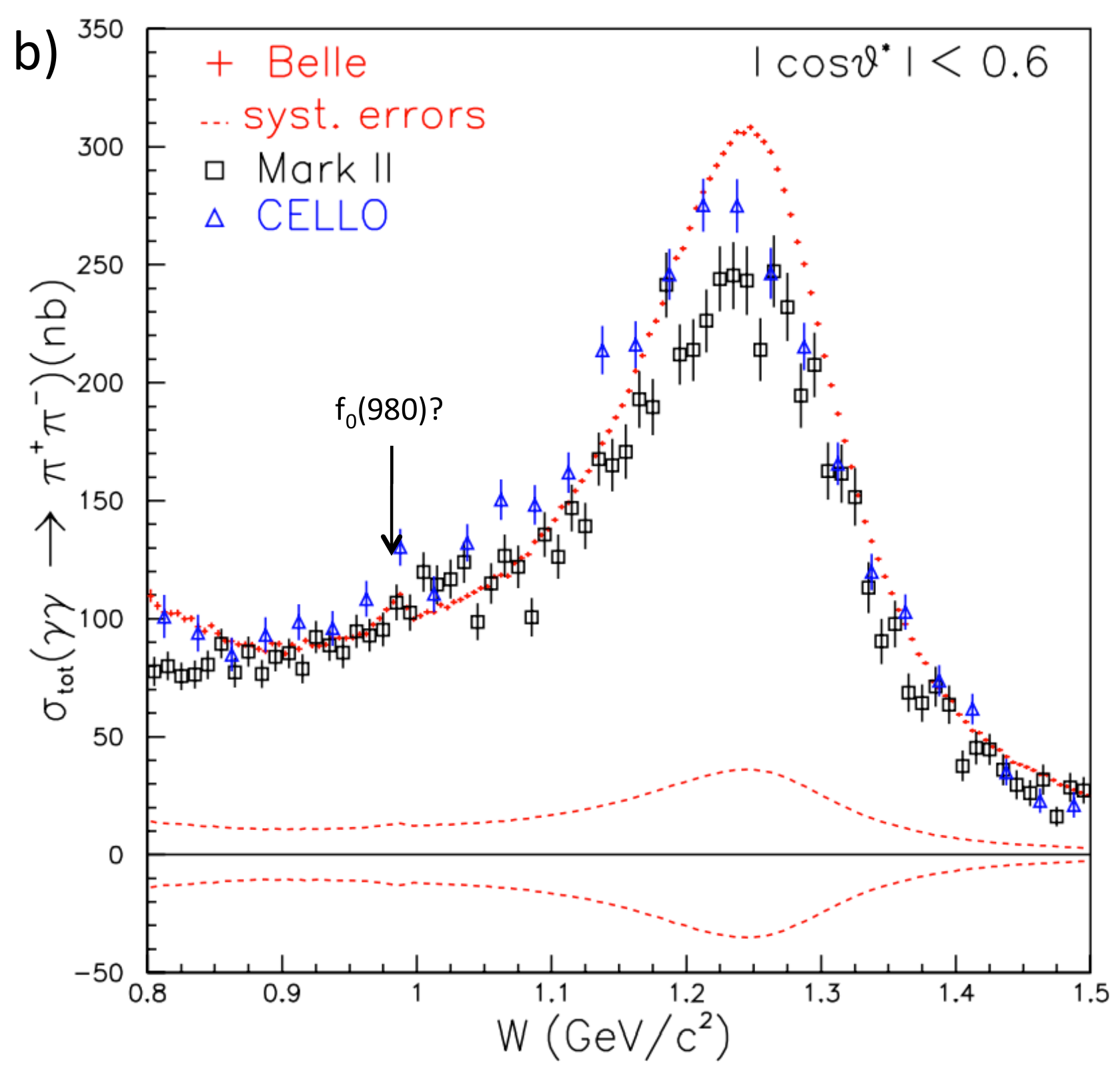}
\end{minipage}
\begin{minipage}[t]{55mm}
  \includegraphics[height=0.8\textwidth,width=0.9\textwidth]{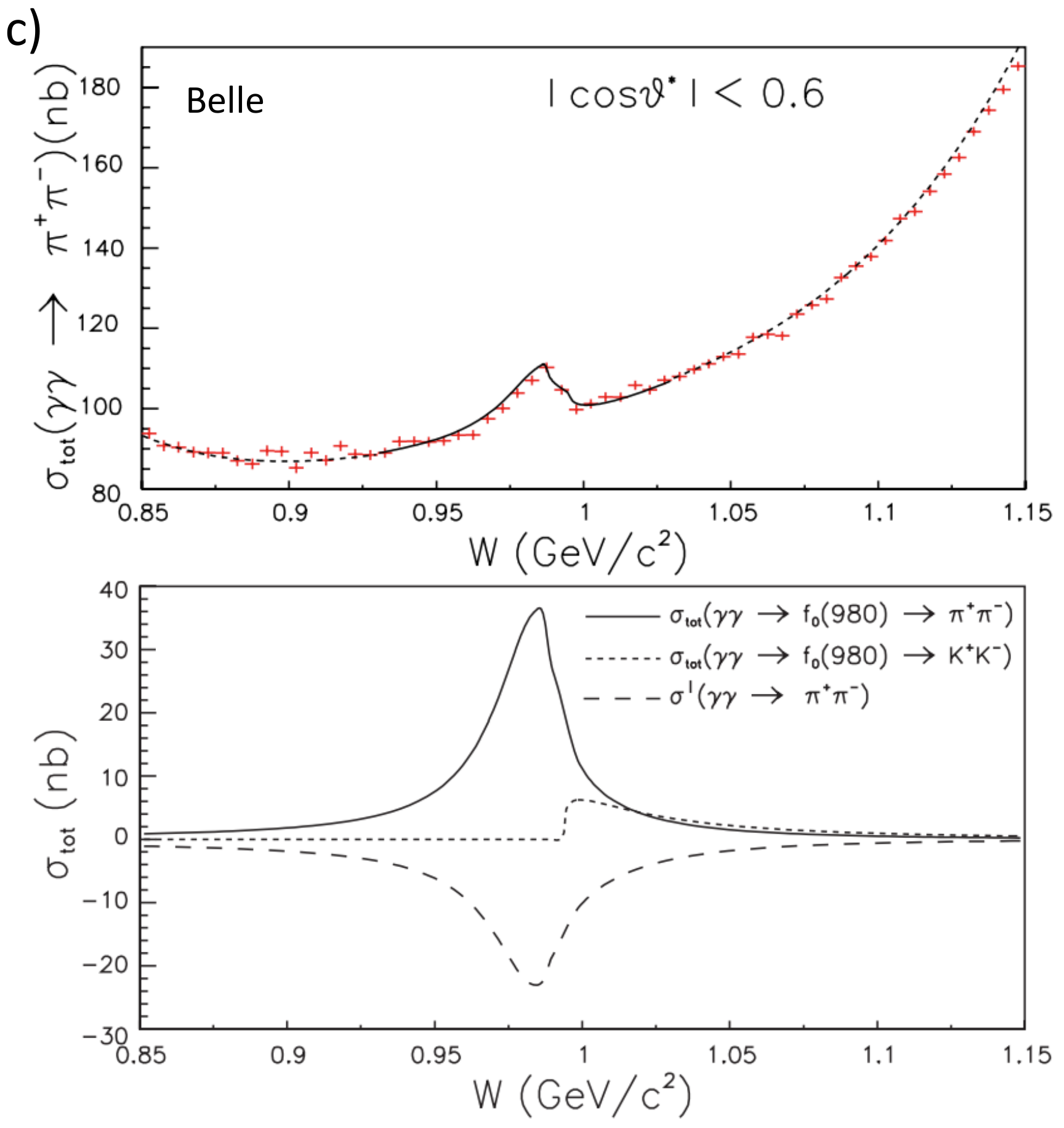}
\end{minipage}
\hspace{\fill}
\caption{ \footnotesize {\bf a)} A cartoon that illustrates the relation between $\gamma\gamma$
production measurements and the internal structure of neutral mesons.
{\bf b)}  $\sigma(\gamma\gamma\rightarrow\pi^+\pi^-$) measurements from Cello (triangles),
Mark~II (open squares) and Belle (red dots). The dashed red line indicates the size of Belle's 
systematic errors {\bf c)}  (upper) An expanded view of the Belle measurements in the vicinity of
the $f_0(980)$ with fit results shown by the curved line.  (lower) The $f_0(980)$ components of the fit:
total $f_0(980)\rightarrow\pi^+\pi^-$ (solid curve); $f_0(980)\rightarrow K\bar{K}$ (short dashes);
effect of $f_0(980)$interference of the non-resonant $\pi^+\pi^-$ background (long dashes). 
}
\label{fig:f0980}
\end{figure}

The measurement of $\gamma\gamma\rightarrow f_0(980) \rightarrow \pi^+\pi^-$ is difficult
with Belle because of a huge background from the QED process $\gamma\gamma\rightarrow \mu^+\mu^-$.
In the $\pi^+\pi^-$ invariant-mass region of interest for this measurement, the pions and
muons have low laboratory momenta and do not reach the Belle muon identification system.
Nevertheless, the different responses of the CsI crystals in Belle's electromagnetic calorimeter
to pions and muons and the huge luminosity of KEKB allow for a mass-bin by mass-bin statistical
separation of the pion and muon contributions.  The blue triangles and black squares in Fig.~\ref{fig:f0980}b
show results from previous measurements by Cello~\cite{cello} and Mark~II~\cite{mark2}, where
there is no sign of any resonance-like behavior in the 980~MeV region.  The small red dots in the
same figure are not a theoretical curve or the results of MC calculations; these are, instead,
Belle measurements with statistical error bars that are about the size as the data points themselves~\cite{belle_f0980}.
The upper panel of Fig.~\ref{fig:f0980}c provides an expanded view of the Belle results near
the $f_0(980)$ mass, where a distinct structure near 980~MeV is evident.  This structure does not have
a simple Breit Wigner line shape because of strong interference with the helicity=0,
non-resonant $\pi^+\pi^-$ background and a distortion caused by the opening of the $f_0(980)\rightarrow K\bar{K}$
at the $2m_K$ threshold.  The $f_0(980)$ is fit with a coherent Flatt\`{e}-like lineshape~\cite{flatte76,achasov05}
using parameters determined by BESII~\cite{bes_f0980}  that takes these effects into account.  The
components of the resulting fit are shown in the lower panel of Fig.~\ref{fig:f0980}c.  The
fit gives an $f_0(980)$ mass and $\pi\pi$ partial width of $M=985.6^{+1.2~~+1.1}_{-1.5~~-1.6}$~MeV and
$\Gamma_{\pi\pi}=34.2 ^{+13.9~~+8.8}_{-11.8~~-2.5}$~MeV and a $\gamma\gamma$ partial width of
$\Gamma_{\gamma\gamma}=205^{+95~~+147}_{-83~~-117}$~eV, where the first errors are statistical and
the second systematic.  The main systematic error on $\Gamma_{\gamma\gamma}$ is from the cross section
normalization that, in turn, is sensitive to the modeling of the non-resonant $\pi^+\pi^-$ background.

Belle also studied $f_0(980)$ production in the $\gamma\gamma\rightarrow f_0(980)\rightarrow\pi^0\pi^0$
channel, where $\mu^+\mu^-$ and non-resonant $\pi^0\pi^0$ backgrounds are not an issue. In Fig.~\ref{fig:f02pi0}a,
Belle results~\cite{belle_f02pi0} for $\sigma(\gamma\gamma\rightarrow\pi^0\pi^0)$ are shown as
red diamonds (with invisible statistical error bars) together with previous results from the
Crystal Ball experiment~\cite{xtalball_f02pi0} shown as black solid circles with error bars. 
Here again the Belle results represent a huge improvement in statistical precision.  The results of Belle
fits to the differential cross section measurements are shown in Fig.~\ref{fig:f02pi0}b, where a
distinct signal for an $S$-wave resonance near 980~MeV is found with mass
$982.2\pm 1.0^{+8.1}_{-8.0}$~MeV and $\Gamma_{\gamma\gamma}(f_0)= 286 \pm 17 ^{+211}_{-70}$~eV; these values
agree well with Belle's results from the $\pi^+\pi^-$ channel but with different sources of systematic
errors.

\begin{figure}[htb]
\begin{minipage}[t]{55mm}
  \includegraphics[height=0.8\textwidth,width=0.9\textwidth]{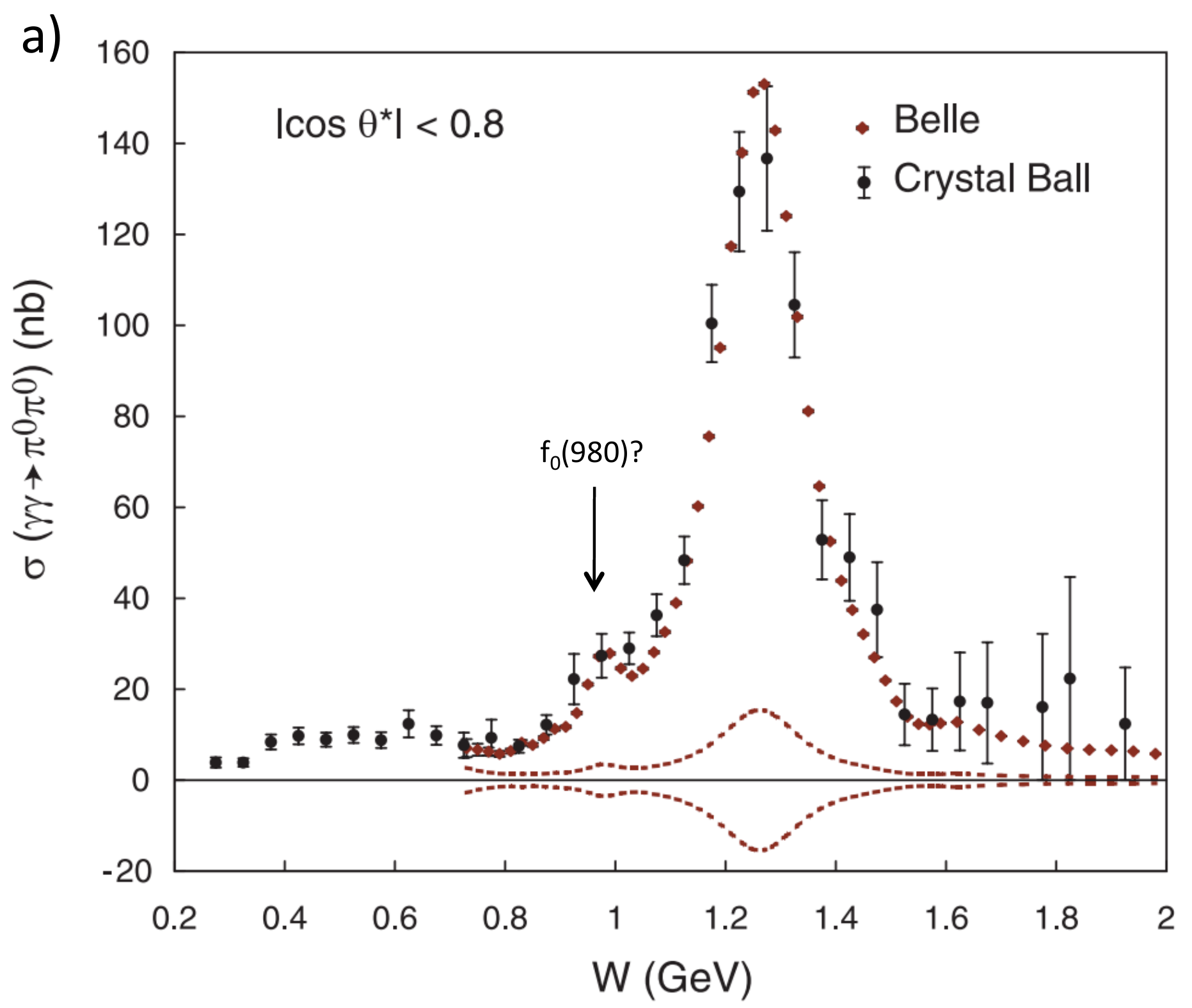}
\end{minipage}
\begin{minipage}[t]{55mm}
  \includegraphics[height=0.8\textwidth,width=0.9\textwidth]{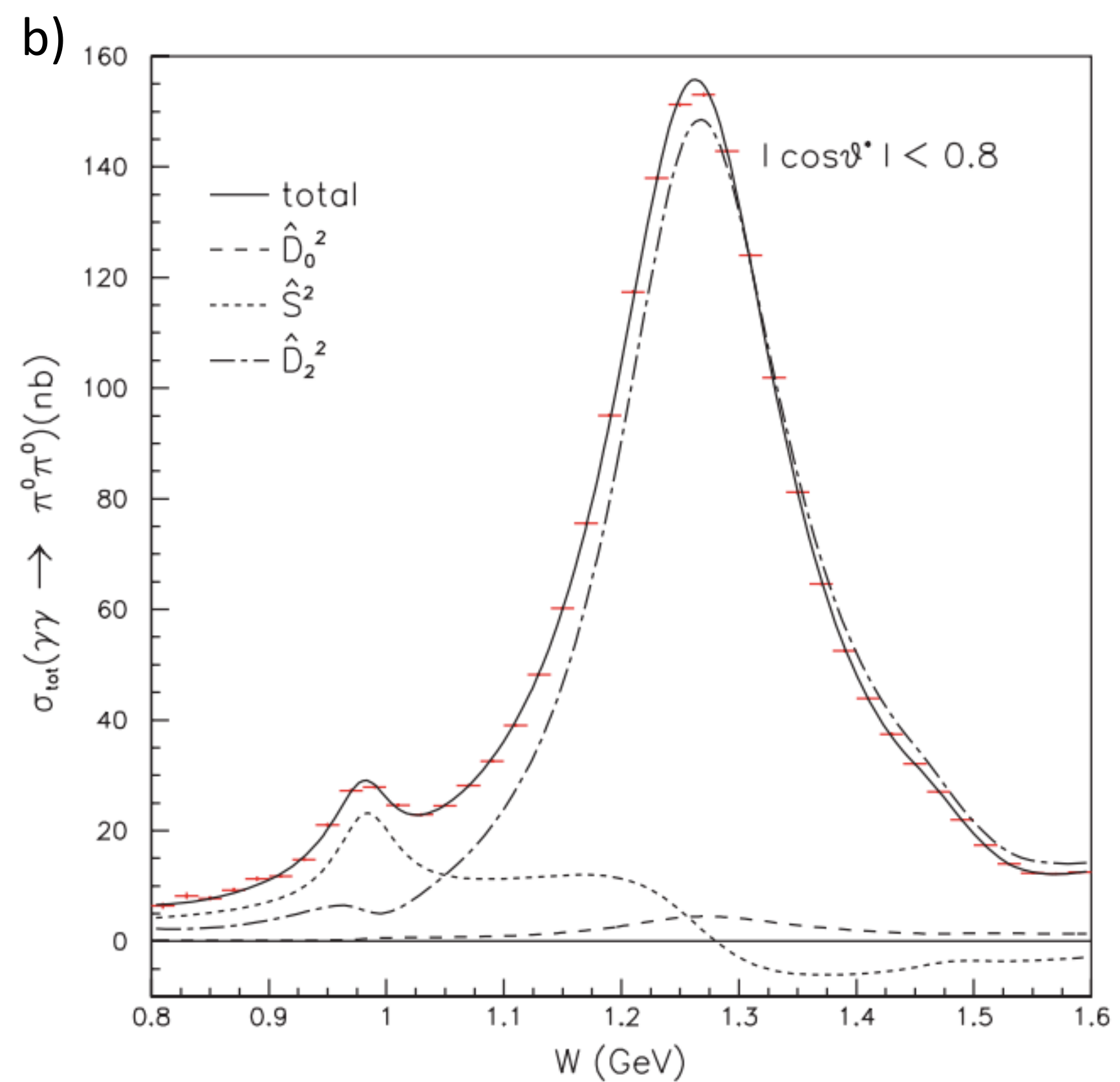}
\end{minipage}
\begin{minipage}[t]{55mm}
  \includegraphics[height=0.8\textwidth,width=0.9\textwidth]{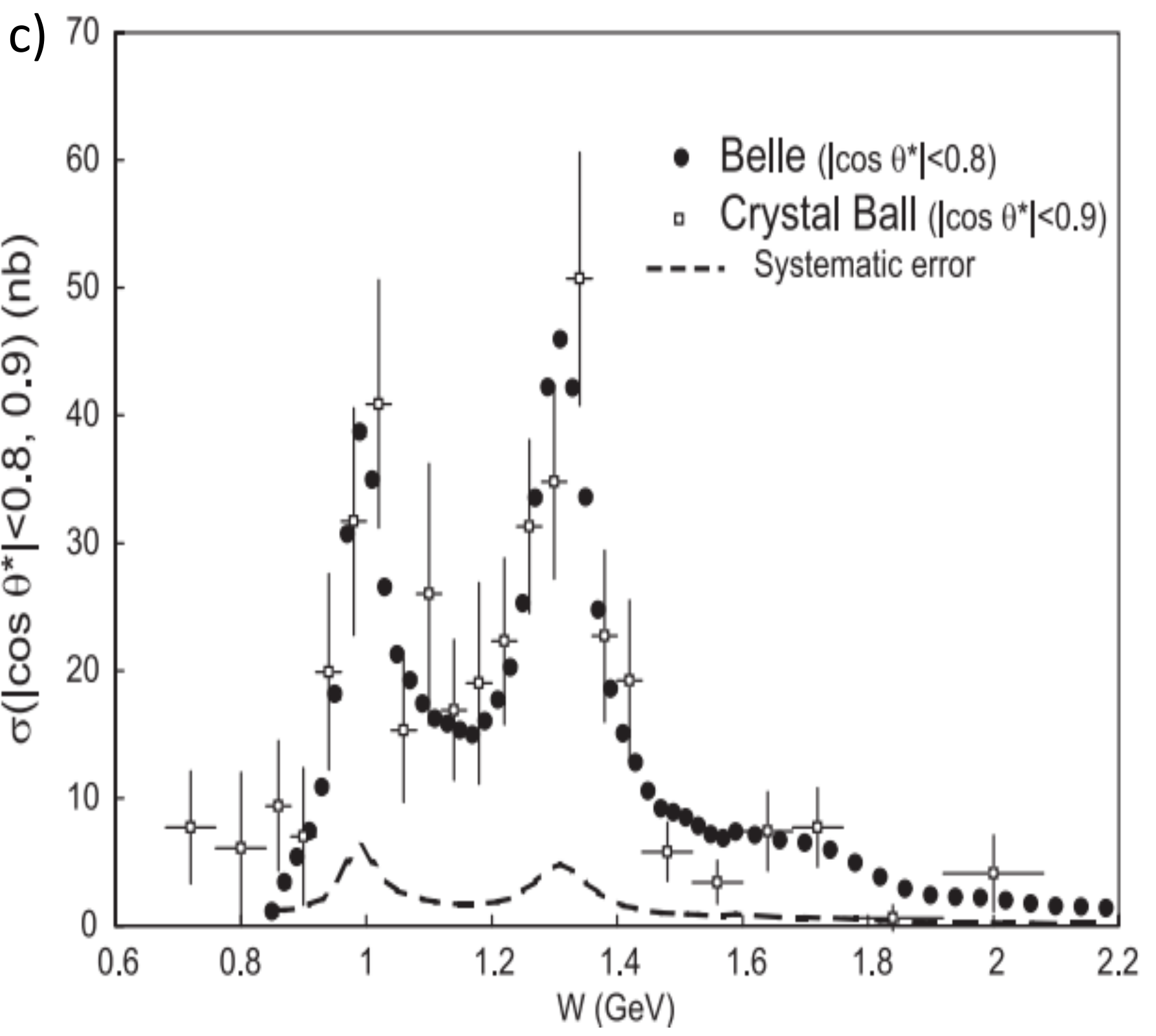}
\end{minipage}
\hspace{\fill}
\caption{ \footnotesize {\bf a)}  Belle (red dots) and Crystal Ball (solid circles) results
for $\sigma(\gamma\gamma\rightarrow \pi^0\pi^0)$.  The dashed line indicates the size of Belle's
systematic errors. {\bf b)}  Belle results for $\sigma(\gamma\gamma\rightarrow\pi^0\pi^0)$ with the results of
the Belle fit: total fit (solid curve); $S$-wave (short dashes); helicity=2 $D$-wave (dash-dot);
helicity=0 $D$-wave (long dashes).
{\bf c)}  Belle $\sigma(\gamma\gamma\rightarrow\eta\pi^0)$ measurements (solid dots) together
with previous Crystal Ball results.  The dashed curve indicates the size of Belle's systematic errors.
}
\label{fig:f02pi0}
\end{figure}

Belle also studied two-photon production of the isovector $a_0^0(980)$ scalar in the 
$\gamma\gamma\rightarrow a_0^0(980)\rightarrow\eta\pi^0$ channel~\cite{belle_a0pi0eta}. Belle's 
$\sigma(\gamma\gamma\rightarrow\eta\pi^0)$ results are shown as black dots in
Fig.~\ref{fig:f02pi0}c with previous measurements from the Crystal Ball
shown as open circles with error bars.~\cite{xtalball_a0pi0eta}.
Belle results agree well with the previous measurements but with substantially improved precision.
The Belle results for the $a_0^0$ mass, total width and $\gamma\gamma$ partial width are:
$M=982.3^{+0.6~~+3.1}_{-0.7~~-4.7}$~MeV; $\Gamma_{\rm tot}=75.6\pm 1.6^{+17.4}_{-10.0}$~MeV; and 
$\Gamma_{\gamma\gamma}\times {\mathcal B}(a_0\rightarrow \eta\pi^0)=128^{+3~~+502}_{-2~~-43}$~eV.
The large positive systematic error on the $\gamma\gamma$ partial width is associated
with uncertain interference effects with higher $\eta\pi^0$ resonances, which were not
considered in previous measurements. 

The Belle $\Gamma_{\gamma\gamma}(f_0)$ results are inconsistent with expectations for a pure $q\bar{q}$
meson and consistent with the four-quark model prediction of 270~eV provided in ref.~\cite{achasov82}.
The impact of Belle results on the understanding of the light scalar mesons is discussed in
refs.~\cite{achasov13} and \cite{pennington14}. 

Belle published ten papers on $\gamma\gamma$ production of six light meson
channels: $\pi^+\pi^-$, $\pi^0\pi^0$, $\eta\pi^0$, $\eta\eta$, $K^+K^-$ and $K_SK_S$. These
papers all include game-changing improvements in statistical precision over previous
work (similar to the examples given above) and include analyses of twenty well identified
meson states that include, usually for the first time, consideration of angular distributions
and the effects of interference.

\subsection{Spin polarimetry for quark jets}

The strongly interacting particles in the SM are quarks and gluons.  The strongly
interacting particles in Nature are hadrons.  Presumably the transition of quarks and gluons into
hadrons is described by long-distance QCD, but calculations of the processes that are involved are
hopelessly complicated.  Attempts to cope with these difficulties by using ``QCD-motivated'' models
have had only modest success.   Usually, the transitions between quarks and hadrons are parameterized
by experimentally measured fragmentation functions $D_q^h(z,p_{h\perp}^2)$, which are probability densities
for a quark of flavor $q$ to produce a hadron $h$ with a fraction $z$ of the quark's original momentum
and with a transverse momentum relative to the quark direction of  $|p_{h\perp}|$, as illustrated in the
upper part of Fig.~\ref{fig:collins-1}a.  Measuring these fragmentation functions is an important
(but unsung) part of the research program of most experiments (see {\it e.g.}, ref.~\cite{belle_frag}).   

\begin{figure}[htb]
\begin{minipage}[t]{87mm}
  \includegraphics[height=0.5\textwidth,width=0.8\textwidth]{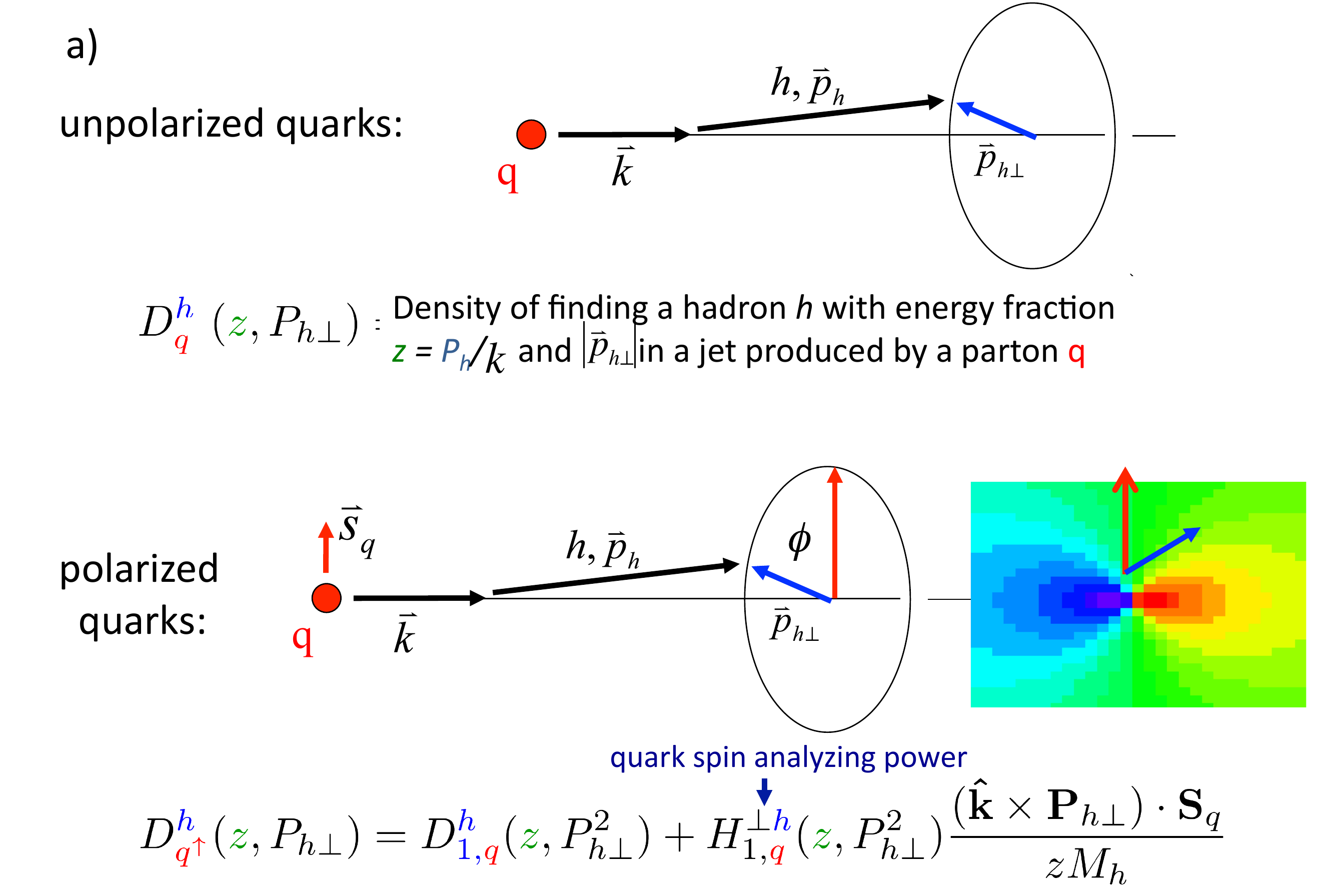}
\end{minipage}
\begin{minipage}[t]{87mm}
  \includegraphics[height=0.5\textwidth,width=0.8\textwidth]{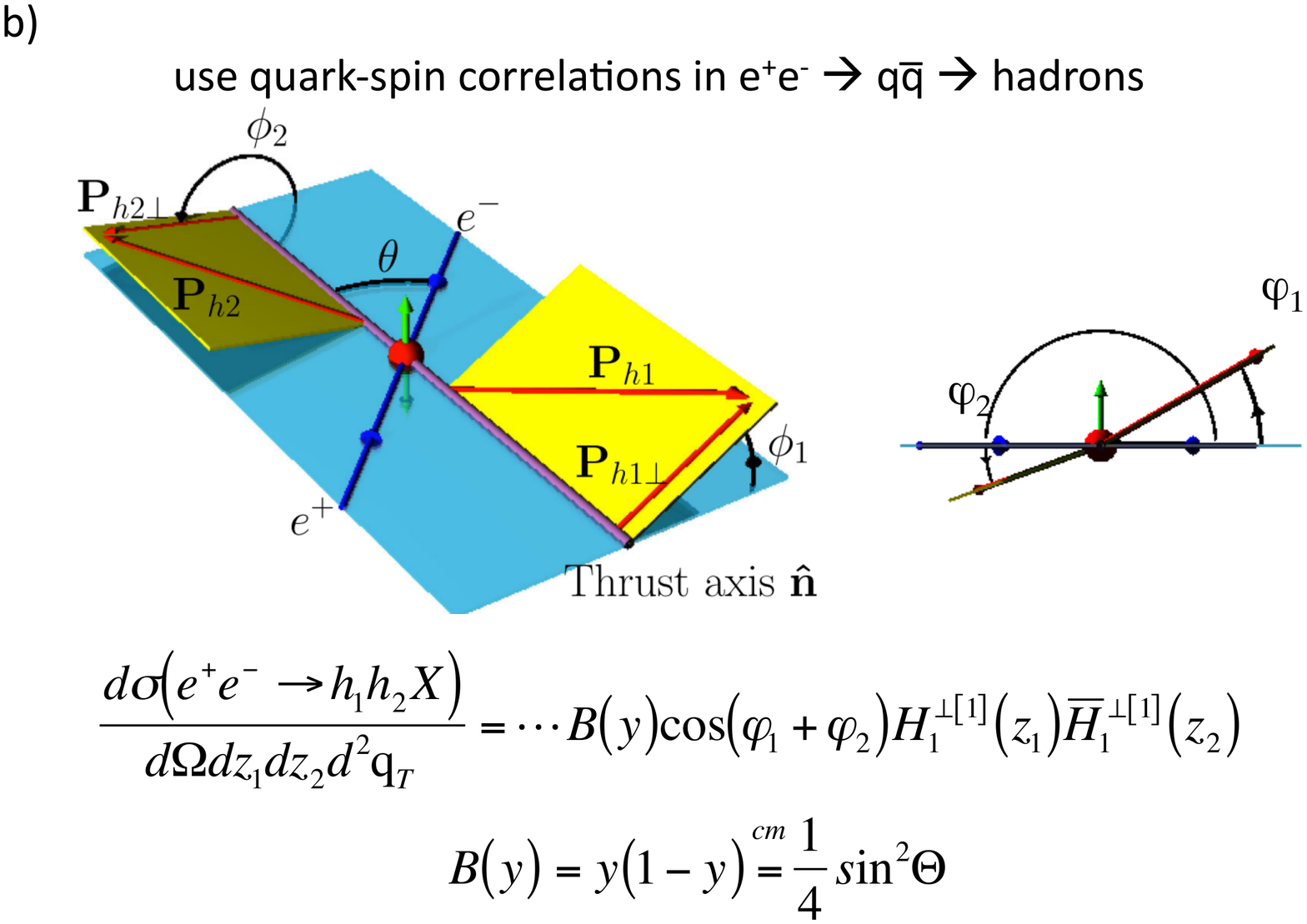}
\end{minipage}
\hspace{\fill}
\caption{ \footnotesize {\bf a)} Illustration of unpolarized (upper) and polarized (lower)
quark fragmentation functions.
{\bf b)}  The principle of measurement of the product of two Collins spin fragmentation
functions using $\ee\rightarrow q\bar{q}$ annihilations.  Here the blue plane is defined
by the thrust axis of the event (purple line) and the incoming $\ee$ direction (blue line).
}
\label{fig:collins-1}
\end{figure}

If the quark is polarized, the fragmentation density can also depend on the azimuthal angle
around the the quark's initial momentum direction as illustrated in the lower part of 
Fig.~\ref{fig:collins-1}a.  This was first discussed by Collins~\cite{collins93}, who introduced a
second term in the fragmentation, $H_{1,q}^{\perp h}(z, p_{h\perp}^2 )$, as a first-order characterization
of this azimuthal modulation.  Thus, if $H_{1,q}^{\perp h}$ is known, the azimuthal distribution of hadrons
$h$ in a jet can be used as a ``polarimeter'' to determine the polarization of its parent quark.

There is growing interest in the proton's transverse spin structure and
some important observables require measurements of quark spin directions~\cite{quark-spin}.
For this, independent determinations of the Collins function are needed, and this requires
sources of quarks with well defined spin orientation.

In $\ee\rightarrow q\bar{q}$ annihilations the individual quarks are not polarized. However,
since the spin of the $q\bar{q}$ system is aligned as either $|J;J_z\rangle = |1;+1\rangle$
or $|1;-1\rangle$, with no $|1;0\rangle$, the spins of the individual $q$ and $\bar{q}$ are
tightly correlated.  Because of this, measurements of the azimuthal angles of pairs of particles from
opposite quark jets can be used to extract products of two Collins functions as illustrated
in Fig.~\ref{fig:collins-1}b~\cite{boer09}.     

Belle used this technique to make first measurements of the Collins function~\cite{belle_collins-1}.
Figure~\ref{fig:collins-2}a shows the the $2\phi_0$ distribution for a typical $(z_1,z_2)$
bin, where a clear $\cos 2\phi_0$ modulation, with an amplitude that is six standard deviations
from zero, is apparent.  Distributions for these modulation amplitudes for ten $(z_1,z_2)$
bins, from ref.~\cite{belle_collins-2} are shown in Fig.~\ref{fig:collins-2}b. Here results
from  492~fb$^{-1}$ data sample accumulated at the $\Upsilon (4S)$ resonance peak (green points)
and a much smaller, 29~fb$^{-1}$ data sample taken at energies below the $\Upsilon(4S)$.  (Since
this analysis is restricted to high thrust events ($T>0.8)$, contamination of the $\Upsilon(4S)$
data sample results from $B$-mesons is negligibly small.)

\begin{figure}[htb]
\begin{minipage}[t]{87mm}
  \includegraphics[height=0.5\textwidth,width=0.8\textwidth]{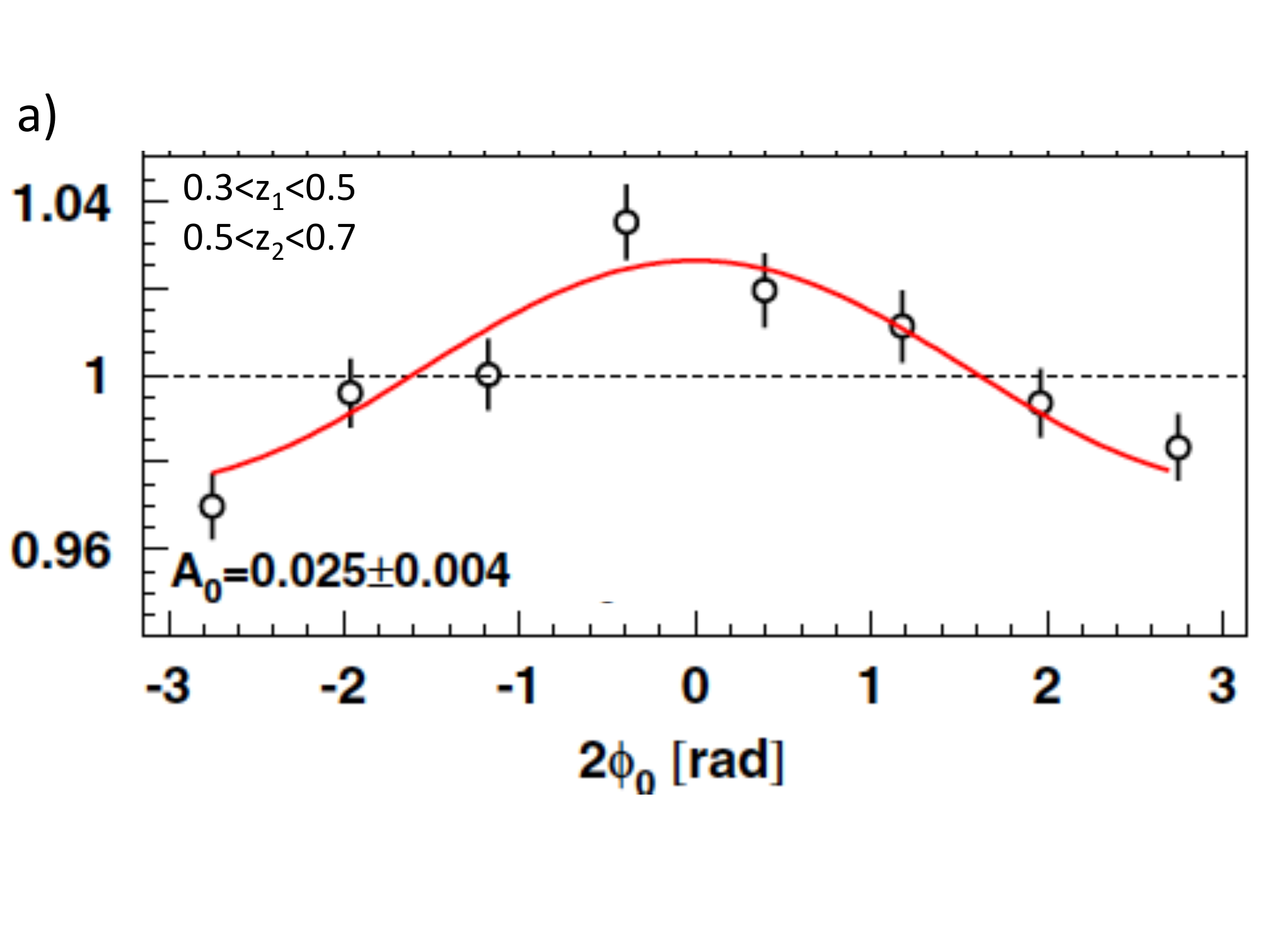}
\end{minipage}
\begin{minipage}[t]{87mm}
  \includegraphics[height=0.5\textwidth,width=0.8\textwidth]{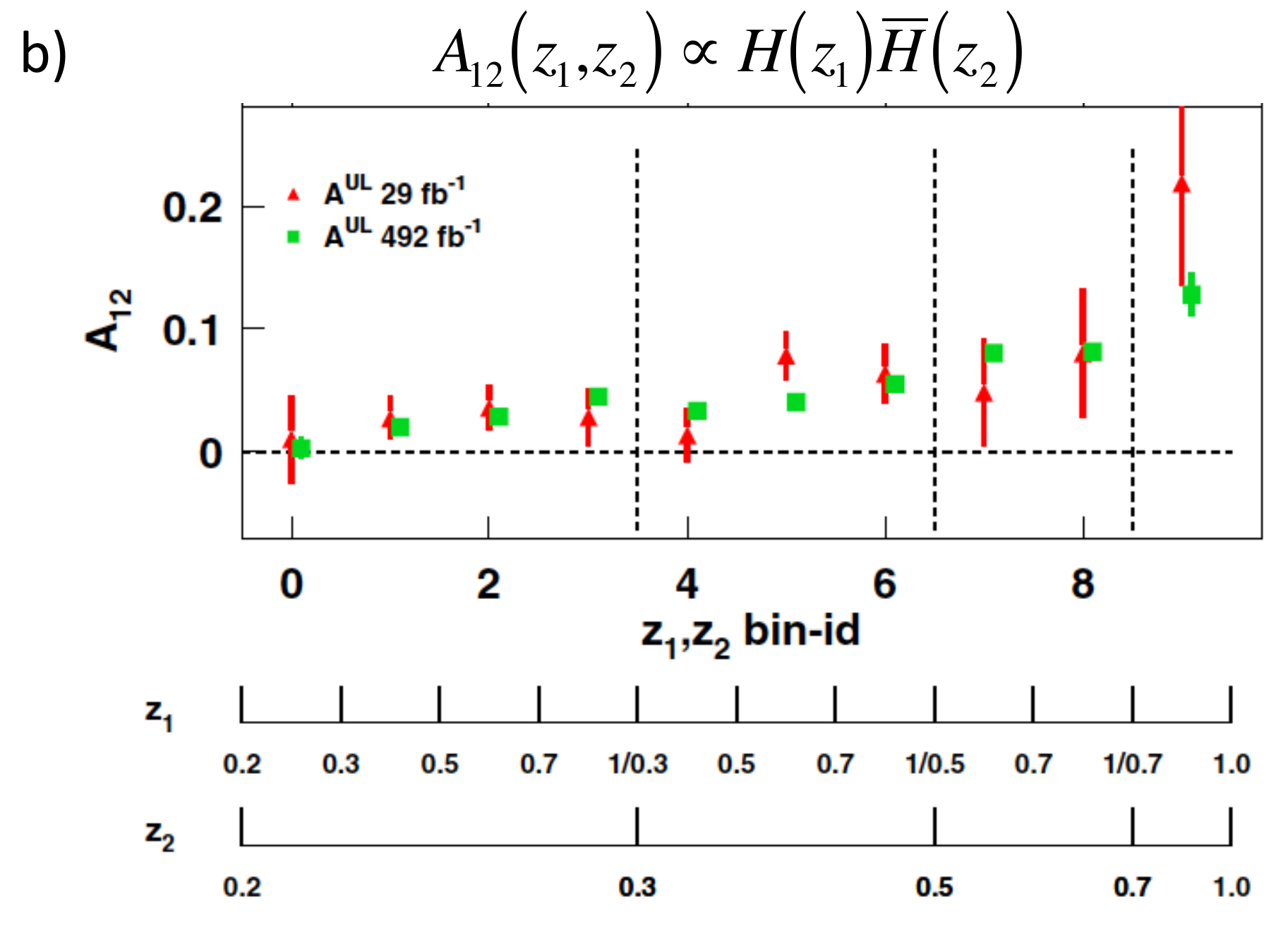}
\end{minipage}
\hspace{\fill}
\caption{ \footnotesize {\bf a)} The $2\phi_0$ distribution for selected events in
the $(z_1,z_2)=(0.4,0.6)$ data bin. Here $\phi_0$ is the azimuthal angle between
the $h_1$ and $h_2$ directions (see ref.~\cite{belle_collins-1}).
{\bf b)} The green points are the measured $\cos(\phi_1 +\phi_2)$ modulation amplitudes
at $\sqrt{s}=10.58$~GeV; the red points are from a small data set taken at a slightly 
lower c.m. energy (from ref.~\cite{belle_collins-2}).
}
\label{fig:collins-2}
\end{figure}

These first measurements of Collins spin fragmentation functions have had a big effect
on spin physics experiments.  Belle refs.~\cite{belle_collins-1}~and~\cite{belle_collins-2}
have been cited 206~and~145 times, respectively.  In addition, they have also stimulated measurements
by the BaBar~\cite{babar_collins} amd BESIII~\cite{bes_collins} experiments.

\section{Summary and Outlook}

In addition to achieving all of its original goals of exploring $CP$ violations in the $B$ meson sector,
Belle has had an interesting and diverse program of investigations that cover a wide variety of subjects. 
While many of these subjects, such as the discoveries of $D^0\leftrightarrow\bar{D}^0$ mixing and $XYZ$ mesons,
are generally well known, there are many others that are less publicized but have had a major impact on their
particular specialized field.  In this presentation, I only had time to cover three of these subjects. 

This broad range of investigation was mostly facilitated by the huge luminosity that was provided by
the KEKB collider, which acheived a world-record-breaking instantaneous luminosity in excess of
$2\times 10^{34}$cm$^{-2}$s$^{-1}$, twice the original design value, and that was generally considered to be
wildly over optimistic when it was first proposed~\cite{quinn00}.  This bodes well for BelleII~\cite{belleii}
and SuperKEKB~\cite{superkekb}.  We can look forward to all sorts of interesting surprises and unexpected
phenomena from this facility.

\section{Afterword}

While preparing this proceedings article I was saddened to learn of the passing of my close friend and esteemed
colleague Susumu Okubo.  Susumu was a brilliant theoretical physicist and one of the pioneers in hadron physics,
with important input into many of
the subjects that I touched on in this summary and memories of him and his deep insights
kept recurring to me as I struggled to digest the subject material into some sensible remarks.    I am
grateful for his friendship and all that he taught me and pray that he now rests in peace.

\vspace{.5cm}
\appendix{\bf Acknowledgments} 

I congratulate the organizers of QFTHEP-2015 on their successful and interesting meeting.
This work was supported by the Institute for Basic Science (Korea) under project
code IBS-R016-D1. 

\bibliographystyle{aipproc}




\end{document}